# You can just review things: A digital ethnography of informal peer review

Jay Patel[1], Joel Chan[2]
University of Maryland – College Park

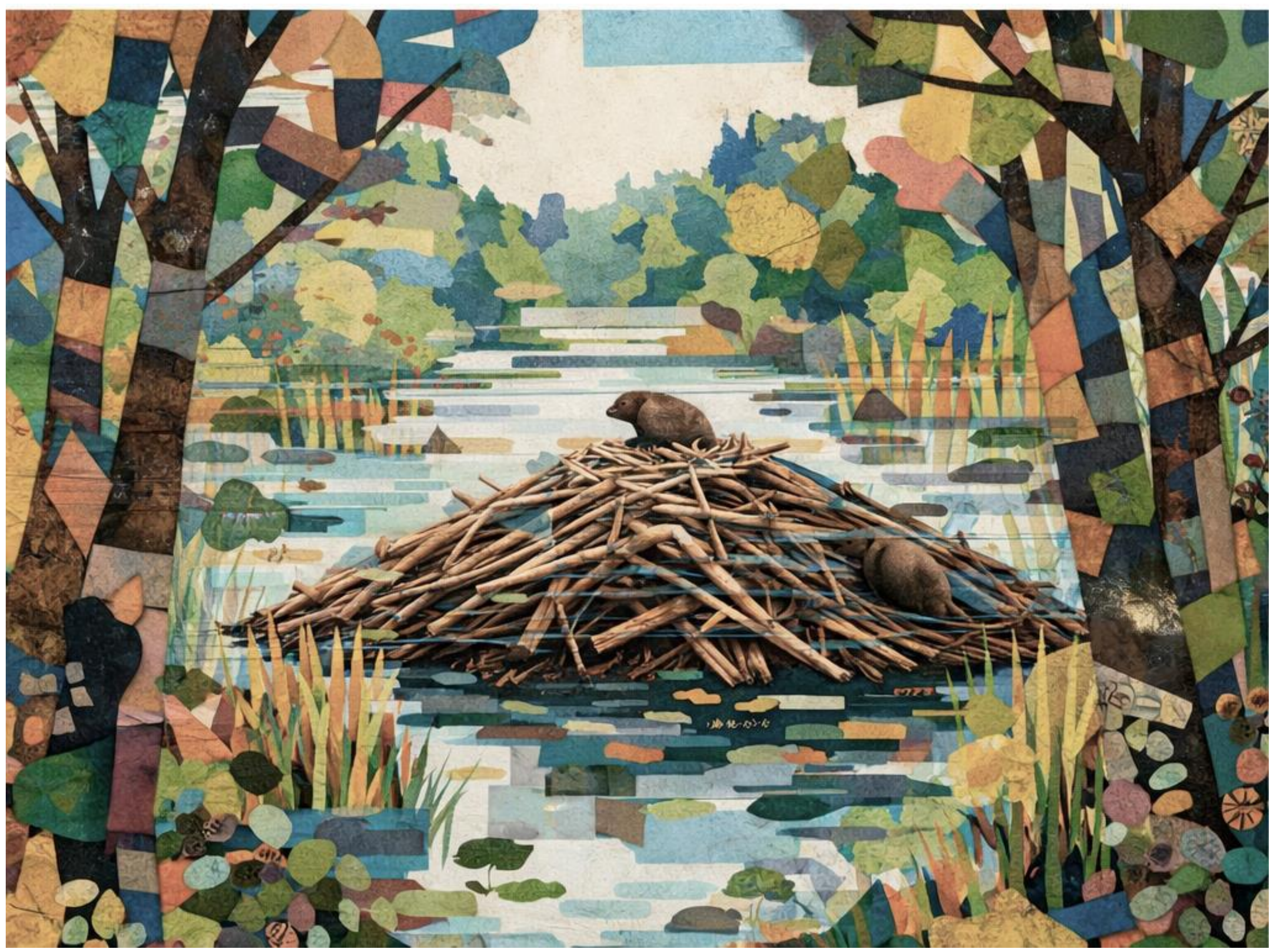

[1] Corresponding author: Address communications to patel93@hey.com, @infotainment.bsky.social, ORCID: https://orcid.org/0000-0003-1040-3607

Hornbake Library
3841 Campus Dr, Office 2118G
College Park, MD 20742
United States

[2] ORCID: https://orcid.org/0000-0003-3000-4160

## Abstract

**Background:** Across scholarly communities, manuscripts face similar evaluative rituals. A learned society or publisher, led by an editor, privately invites reviewers within a specialty to appraise a manuscript's merits through a loosely structured evaluation closed to outsiders, mediated by publishers, and conducted on private platforms. Let's call this *formal peer review*.

**Construct:** Though formal peer review is the norm, scholars are increasingly congregating online to make sense of research reports. These newer and understudied patterns of peer review are often open to outsiders, minimally or unmediated by publishers, and distributed across public platforms. We call this practice, a blend of three open peer review variants, *informal peer review.* Those who practice it, *informal peer reviewers*, may be occasional error detectors or veteran sleuths who surface overlooked issues like plagiarism, fraud, errors, undisclosed conflicts of interest, and a litany of conceptual confusions restricting claims' inferential reach.

Informal peer reviewers also make sense of research decisions, explain jargon, critique clarity, judge value, and connect to related literature. Unlike formal peer reviewers, informal reviewers lack the official authority to make accept-or-reject decisions or sufficient resources to power their activities.

**Questions:** We asked: (1) Who are informal peer reviewers? (2) Where do informal peer reviewers work? (3) How do informal peer reviewers evaluate scholarly reports? (4) What are the impacts of informal reviews on publishers, study authors, and others?

**Methodology:** To answer our questions, we conducted a cross-platform digital ethnography with active participant observation. We traced discourse about research reports flowing from publishers to a network of digital communities and back by immersing ourselves in the communities over four months. After nine and twelve months, we followed-up on all cases to assess long-term outcomes. From a set of fifteen digital communities including social media sites and review platforms, we used theoretical sampling to select twelve cases mentions (ten unique cases, two described twice) and eight meta-commentaries by 26 reviewers for deeper analyses.

**Analyses & Results**: We analyzed our data using open and axial coding, creating 1,080 codes from which we constructed four high-level themes: (1) informal peer reviewers are a diverse *motley crew*, (2) they self-organize their work by nesting *hacked homes* in subpar settings and bridging communities, (3) use deep reviewing strategies uncommon during formal peer review that we call *intensive infrastructural inversion*, and (4) battle resistance akin to *immunologic self-preservation* from the triad of authors, publishers, and editors so that they may be taken seriously.

**Conceptual Contributions:** By abstracting patterns across cases, we concluded that informal peer review (1) is a suitable name and (2) relies on a fragile network of people, platforms, and reviewing practices that we theorize as a minimally governed *patchwork evidence infrastructure*.

**Practical Contributions:** To scale informal peer review, one might extend this patchwork with connective tools, communities, and training programs. Alternatively, one might build bespoke tools and communities. This latter path is possible, though difficult. Those who undertake it should help mature the patchwork infrastructure alongside scholarly values, ease frictions, and reward attempts to extend the scholarly dialogue.

# 1 Introduction

PRE-PUBLICATION PEER REVIEW

> "Comments and criticism are welcome on the intro draft… proposing and testing this hypothesis… Not yet a preprint… (Mõttus, 2024)"

> "Our preprint is out on bioRxiv!... Feedback welcome via GitHub Issues... (Oakden-Rayner, 2018)"

> "I wanted to finish talking to the journal, but the media articles keep coming and even @ylecun is repeating the claims... time for some #openpeerreview! (Furudate, 2018)"

FORMAL PEER REVIEW

> "Shall we call traditional peer review… secret, only involving a couple of reviewers who may not between them have all the required skills, knowledge and experience initial peer review: IPR? (Berkley, 2026a)"

POST-PUBLICATION PEER REVIEW

> "Authors sometimes naively think that once the paper is published, the discussion is over. (Thorp & Phelan, 2025)"

> "…publication is the beginning of peer review (McElreath, 2024)"

> "the real potential of the electronic medium is in accelerating the INTERACTIVE phases of scholarly research... which are really on a continuum… one that has no end… (Harnad, 1992)"

Of the varied forms of peer reviewing practices, one form has been overlooked and withheld its due. It is a neglected middle child crying for unmet attention and respect. Even supposing careful study, its workings and impacts are likely to be misconstrued by confusion with more common variants. Here, we describe the type of peer review that has held our continued fascination in cases that only graze the richness of this phenomenon.

The first case, an early encounter with the *credibility* of scientific diagrams, began on the morning of February 15, 2024, when scholars in the life sciences shared a link to a peer-reviewed paper from the journal *Frontiers in Cellular and Developmental Biology* with their social media followers. This paper could be critiqued on multiple dimensions, though the target-of-interest was an AI-generated image of a rat with abnormally large genitals and misspelled annotations of anatomical features. Other diagrams of cellular signaling pathways, though less comical to non-specialists, were also incorrect and seemed AI-generated to specialists and laypeople. These scientific flaws bulged so egregiously in the paper that scholars wondered to their followers on X, Bluesky, and personal blogs, "...how did Figure 1 get past a peer reviewer?!" (Frontiers [@FrontiersIn], 2024).

Later that day, the journal issued an expression of concern and, soon after, officially retracted the article (Frontiers Editorial, 2024) due to credibility concerns. In their reply to social media discussions, they praised the unconventional route to correction: "Thanks to the crowdsourcing dynamic of open science, we promptly acted upon the community feedback on the AI-generated figures... (Frontiers, 2024)."

In a second and more subtle case, the *connectivity* of economics research to prior work was examined and found to be in jeopardy. When scholars stumbled upon research disconnected from relevant literature, they seem particularly passionate about voicing their disapproval.[3] In a Twitter thread, an economist unraveled the missing information in a paper about the intergenerational effects of slavery and Jim Crow in the 19th-century US (Backhaus, 2024). The co-authors of the paper “Jim Crow and Black Economic Progress after Slavery” (Althoff & Reichardt, 2024) cited a prior paper by the economist Bruce Sacerdote titled "Slavery and the Intergenerational Transmission of Human Capital" (Sacerdote, 2005) in passing and failed to describe its relevant empirical contributions. Althoff and Reichardt, 2024 omitted analyzing relevant data and discussions that were published by Sacerdote 2005. This missing connectivity of data, analyses, and interpretations between related papers seemed like an elitist cold shoulder to prior work, "very common within the Top5 circle, more common than in lower-ranked journals… (Backhaus, 2024)." To date, this oversight remains uncorrected in the published article, and we haven't learned of any signs that the co-authors of the paper are aware of the critique.

[3] Systematic review and meta-analyses, we have found from scientific sleuths, are often disconnected from relevant literature and hence skewed towards erroneous conclusions.

Third, we were alerted to a case of broken *care* expectations on Reddit. After obtaining IRB approval from their university, a UK-based research team scraped public data from r/schizophrenia users discussing their experiences with mental illness (Aksenfeld, 2025). They attempted to anonymize quotes with subtle edits and considered public posts to be available for research projects; the authors published their content analysis in *Current Psychology* on June 7, 2024, remarking that gathering informed consent from many users would be infeasible and that the net benefit of their study would be positive (Lyons et al., 2025). Users on Twitter and r/schizophrenia quickly discovered the paper and voiced their strong disapproval about being studied to the subreddit's moderator (Empty_Insight, 2024), who in turn contacted the lead author of the schizophrenia paper. The authors admitted that despite their successful application of privacy safeguards, they violated a "safe space" and agreed to a retraction. This retraction was accompanied by the publication of a retraction notice on December 18, 2024 (Lyons et al., 2025). Volunteered retractions like these are difficult to find, especially a few months after publication. We surmise that the ethical complications of a sensitive medical topic can speed up scholarly revisions and authors may be motivated to save face.

## 1.1 Gaps and Opportunities

So what are these users doing? One might reply *commenting*, which we define as the practice of discussing a topic. It captures behaviors like praise and offhand remarks that we are not concerned about here. Instead, we are interested in the subcategory of *critical comments that evaluate specific research reports*. Such comments vary in their politeness from neutral to demeaning. Reviews and content analyses of research commentary on social media have described users across social media sites and wikis, but mention critiques only briefly without capturing their potency (Sugimoto et al., 2017; Tsou et al., 2015). See Table 1 for examples of commenting by sentiment.

**Table 1.** Complimentary, neutral, and critical posts by Bluesky user @jbakcoleman.bsky.social.

| **Sentiment** | **Example Post** | |
|---|---|---|
| | specified research report | unspecified research report |
| complimentary | "It's wildly elegant mixture of theory and empirics... Paper of the year."<br>"It's the platonic ideal of a scientific paper, and it landed appropriately in science." | "Reviewed a lovely little paper tackling a big question with little more than counting. Truly nothing I love more than using the least math and statistics possible to answer a question." |
| neutral | "Here's [a] relevant paper on #metascience lit from 2018... This hasn't been hypothetical or hard to see." | not available |
| critical | "This paper claims to provide evidence... but winds up mistaking noise for signal and seems to forget that some folks online are quite influential." | "Managed to get my first ai slop piece to review today. What a giant waste of time."<br>"Pro tip: check the code for emoji" |

Users in the three cases above were also not engaged in science communication *per se*. Instead, they are engaging in the more specific act of evaluating or *reviewing* scholarly reports like talks, posters, papers, or books. Here, we use the sense of reviewing meaning "to go over or examine critically or deliberately (Merriam-Webster, 2025)."

Science communicators, we have found, generally do not critique the study reports that they discuss. Very often, they adopt the ideas at face value and at times do not even link to primary sources. In contrast, our study attends only to cases in which individuals challenged scholarly reports' credibility, creativity, clarity, connectivity, and care.

Reviews like the three cases above were scholar-organized instead of publisher-organized, unsolicited by official agents, intrinsically motivated, and open for public participation. Hence, they attracted many diverse reviewers. At times, these reviewers were distributed across platforms. This seemed to us a clear contrast to the traditional practice of peer reviewing for conferences and journals. Because this form of review is structured and operates under official, widely accepted ways, we call it *formal peer review*[4]: the practice of evaluating research with structured, socially sanctioned forms, rules, tools, and behaviors.[5] It seemed clear that our cases above were *not* formal peer reviews. But what, then, are they? We initially considered whether this form of reviewing could be described as a specific subtype of *post-publication peer review*, but found it difficult to map it to the multiple, at times mutually exclusive definitions (Woods et al., 2022).

The term *open peer review* also seemed broader than our focal phenomenon, subsuming seven distinct practices, including transparent disclosure of reviewers' identities, access to peer review reports, public participation from a broad pool of potential reviewers, direct interactions between author(s) and reviewers or directly between reviewers, posting manuscripts on preprint servers, final-version commenting, and open platforms aka "decoupled review" (Ross-Hellauer, 2017). This conceptual uncertainty was a key motivation for our study. As will be clear later in this paper, our digital ethnography suggests that the existing term "informal peer review" is a fitting name for the phenomenon, as the property of being scholar-organized rather than publisher-organized seems to be a central distinguishing feature of the phenomenon.[6]

---

[4] In the literature, the term "traditional peer review" is used more often, though we wish to make a more precise contrast between formality and informality inspired by our data.

[5] One definition of formal peer review that we favor is inclusive, extending its scope to "any procedure in science that is used to allocate scarce resources by invoking expert judgment on the epistemic qualities of an object" (Reinhart & Schendzielorz, 2024). This widened view is acceptable, according to the authors, because it allows scholars to relate research on diverse forms of peer review, surface more abstract insights, and it accords with the historical trend of an evolving construct. In an editorial, a mathematician commenting on the value of preprint review similarly proffered to "use the phrase "formal peer review" for the kind that is organized by a journal and "informal peer review" for the less official scrutiny that is carried out whenever an academic reads an article and judges it. (Gowers, 2017)

[6] We debated alternative names like public reviews, independent reviews, many reviews, modern peer review, democratic peer review, journal-independent review, unbundled review, grassroots peer review,

This conceptual tangle was complemented by a theory vacuum. We did not find any explicit, data-driven descriptive theories of peer review[7] (formal or informal, explanatory or predictive) that summarize empirical regularities and uncover the mechanisms of successful and unsuccessful reviews that could satisfactorily explain what we saw in the cases of informal review. For instance, Reinhart and Schendzielorz, 2024 were motivated by the call to theorize peer review in Hug, 2022 and have characterized peer review as a form of *governance* akin to a democracy able to self-regulate. This theorization seemed to capture formal peer review, where scholars elect reviewers using nomination processes regulated by publishers who leverage specialized skillsets for the collective good. But how, if at all, might this governance theory of peer review apply to informal review? Does informal review legitimize its decisions somehow? If so, do they mirror the legitimizing decisions of formal peer reviews' manuscript accept-and-reject decisions?

Metascience does not even fully understand informal review at the level of *empirical evidence*[8]. Only a few empirical studies have analyzed cases of informal reviews. The empirical literature so far, includes case studies and some coarse statistical analyses of large datasets. Wakeling et al. 2020, studying commentaries on PLOS articles, bemoaned "how little publicly available empirical evidence" exists and that we need "further work to understand how such innovations can be more seamlessly embedded into researcher workflows."

---

community review, swarm critical appraisal, organic peer review, organic critical appraisal, crowdsourced peer review, and sensemaking. One scholar critiquing the small teams, secretive nature, lack of payment, and homogeneity of expertise in formal peer review procedures facetiously proposed naming it "initial peer review (IPR)" (Berkley, 2026a). Ultimately, we settled on informal review for its simplicity and accuracy. As we discriminated between related constructs and expanded our understanding of peer review, we realized an unmet need to organize them systematically into an ontology. We favor informal peer review's neutrality, its ability to describe the key properties of the phenomenon, its respect for authors of prior empirical studies who used the term, and it is clear contrast with formal peer review.

[7] Our brief literature search using the key terms "theory of peer review" (40 hits) in Google Scholar did not uncover other substantive papers beyond those cited above. A search on the AI-assisted search engine Elicit for "theory of peer review" did not yield new results concerning formal and informal peer review during August 4th, 2025. We ignored mentions of grant peer review due to differing dynamics and purposes. A deeper, cross-disciplinary search mindful of differing jargon and tangential literature may be fruitful.

[8] Using search terms like "open peer review", "post-publication peer review" and common research methods like "interviews", "content analysis", and "ethnography", we failed to unearth many relevant empirical papers.

Finally, we considered that these conceptual, theoretical, and empirical gaps were complemented by a design and practice gap. We recognized in our initial cases an opportunity for a new infrastructure for research evaluation that could support or perhaps replace formal peer review. But we were also cognizant of a large body of investments into practices like this — both institutional projects like PubMed Commons and startups like Peerreview.io (PubMed, 2018, Bingham, 2024) Both were unsuccessful at developing informal review to match the scale and functioning of formal peer review (Heard, 2024; PubMed Commons Archives, 2018). If successful design principles for building informal review tools and communities existed, they would be broadly used and visible in scholarly practice. Principles, guidelines, and communities would be conveniently available to help build and evolve reviewing platforms. But we were unable to find such principles or guidelines, and were left to wonder what might make social media sites and academic forums successful at eliciting informal reviews, where prior efforts to develop these practices had not.

In this paper, we report insights from our digital ethnography of informal peer review. We discovered significant conceptual, theoretical, empirical, and design uncertainties around this emerging form of scholarly review. To encourage deeper understanding and extensions of our work, we authored this report using the detailed values-based reporting items from the Big Q Qualitative Reporting Guidelines (BQQRG) and the APA JARS-Qualitative Reporting Guidelines (Braun & Clarke, 2024; Qualitative Research Design, 2024). We recommend reading both guidelines fully.

### 1.2 Research Questions

Broadly, we aimed to understand the nature of informal peer review relative to formal peer review and inspire solutions for reforming inefficiencies in both practices. Our grounding in the governance theory of peer review by Reinhart & Schendzielorz, 2024, structured our inquiry around reviewing setting, behavior, and functions. These components subsume our four research questions:

Setting: reviewers and settings

RQ1: Who are informal reviewers?

RQ2: Where do informal reviewers work?

Behavior: methods and tools

RQ3: How do informal reviewers evaluate scholarly reports?

Functions: impacts and failures

RQ4: What are the impacts of informal reviews?

Answering these four questions is of interest to metascientists, science and technology studies (STS) researchers, and scholarly publishers who seek to understand the production and evaluation of knowledge systemically. Tool-builders with ambitions to harness the wisdom of the crowd may use the insights below to guide the design and development of informal review technologies.

## 2 Positionality and Reflexivity

I (JP) immersed myself in the informal review community as a third-year Ph.D. student with the oversight of two faculty mentors (IP, JC) and approached this ethnography with a deep background in academia spanning graduate education in the social, behavioral, and educational sciences. I arrived with a decade of experience doing research across six labs, working in industry research roles, and coordinating independent metascientific and science software projects. Although I've immersed myself in quantitative and qualitative research methods, a background in descriptive and inferential statistics, and theorizing about natural and social phenomena, I acknowledge that some cases of informal reviews required me to do further research and trust that claims agreed upon by academic experts in a given domain were indeed accurate. I did *not* seek to validate the critiques in informal review cases definitively, but study and report them. My broad academic experiences enabled me to comfortably traverse disciplines and source cases for analysis without facing paralyzing knowledge barriers.

In the early phase of my graduate education and research career, I learned and implemented study design, quantitative and multi-method data analysis, writing, and peer reviewing from graduate advisors and conference presenters. Later, I was inspired by the open science movement and the associated blossoming of metascientific research. This, coupled with personal experiences conducting some unreliable research for a PI in a prior Ph.D. program, led me to leave graduate school for several years in search of work and the independence to pursue personal projects.

I do not have any ties that I interpret as significant financial or social conflicts of interest, though I participate in science reform communities and initiatives with an emphasis on reforming peer review and evidence synthesis. My employment at my university and consulting across employment sectors are independent of this study. Personal experiences with low- and high-quality research, often in the same research team, are most relevant to this study. I have been driven by a need to reveal the well-documented examples of low-quality, redundant research and to simply report the easiest cases to sample. As such, I focused on informal reviews that surface problems with research quality and impact. Over time, I noticed that informal reviews are much more complex than the identification of low-quality research. They span a continuum of importance and attend to diverse features of scholarly reports.

I (JP) acknowledge my strong bias *against* formal peer review and in favor of more informal options. These biases have formed from a decade of cross-disciplinary readings, private musings, social experiences, conference activities, and personal encounters with a broader menu of reviewing options online. The insights from this study also imply that reforms are needed. My biases are ineluctable and even valuable, so I will not position myself with a faux neutrality that promises unmerited impartiality. I reject the possibility of complete impartiality and appropriate distance from the research, a lesson from the *symmetry principle*. Instead, I subscribe to a stance that permits biases if they are acknowledged and shown to be useful – *weak asymmetry* (Pels, 1996). My biases have impelled me to undertake this study, investigate common hesitations around informal reviewing, and network with tool-builders to help them recognize the benefits and imperfections of informal reviewing. I also acknowledge, as some data out of this study’s scope shows, that informal peer review can err, failing spectacularly in some cases. Given the length of this report, I reserve deeper elaboration about the ethics and etiquette of informal peer review for an upcoming paper based on a multi-year, multi-study collaborative research program.

Although I did not engage in reflexive journaling in my memos formally, I reflected on how my past shaped my data sourcing and perspectives on research reform in team discussions with faculty mentor and co-author JC. I've followed some of the informal reviewers described in the Methodology and Insights sections for years before the study was designed. Only a few of them are familiar to me from in-person metascientific and open science conferences. The rest have lingered in the periphery of my social media feeds, surfacing as I provided social media algorithms signals to recommend more relevant content. Within an academic year, I transformed from a casual observer or outsider to an insider with long-term interests in studying and supporting informal reviewing.

# 3 Literature Review

To understand the functions and sociopolitical context of reviewing, we must trace its misunderstood and understudied history. Let's first consider the evolution of the dominant practice, formal peer review, so that we may later contrast it with informal peer review.

## 3.1 Younger than you look: A brief history of formal peer review

Formal peer review's prehistory, which has been closely investigated by sociologists, historians, and philosophers, begins with the mythologized English editor Henry Oldenburg and his commentaries on work in the journal *Proceedings of the Royal Society*. It also originates from the philosopher of science William Whewell and his 19th-century summary reports in the same journal (Csiszar, 2016, Baldwin, 2018). Like Oldenburg, Whewell's summaries, or *referee reports*, were not intended as quality control filters to gatekeep papers based on judgments of their credibility. They were used, instead, for *communicating* scientific research to laypeople favorably much as science news magazines and websites do today (Baldwin, 2018).

Contrary to Whewell's initial wishes, his casual reports eventually morphed into private referee reports to evaluate manuscripts' quality (Harington, 2018). From then to the mid-20th century, refereeing expanded as scientific judgment on its use diverged. This practice was at first exclusively English, not diffusing to America until the early 20th century. Then, the sudden boom in journal submissions mired editors in manuscripts and the increased governmental investments in research obliged scientists to justify spending millions of taxpayer dollars during an economic downturn (Merriman, 2021). Editorial burden and public accountability for spending taxpayers' money required deep reforms, so scientists shielded themselves under the aegis of peer review blessed by their expert colleagues. Administrators at US National Science Foundation (NSF) used grant peer review with external experts called *referees*, inviting them to review grant applications with an administrator. This new funding distribution model (1) self-regulated funding decisions within the scholarly community, (2) built consensus about proposals, and (3) ensured public accountability of money used (Csiszar, 2016).[9]

In parallel, elite journals like *Science, New England Journal of Medicine, Nature, The Lancet, Physical Review,* and *Philosophical Magazine* recognized grant peer review's value and embraced referees. In their reflections on this modern turn towards *formal* peer review and its use of anonymous and external experts, modern historians have concluded that our current fixations on formal peer review derive from these recent reforms (Csiszar, 2016; Baldwin, 2018). The very "shift from 'refereeing to 'peer review'…established a narrow range of acceptable reviewers and implicitly deemed those without a scientific background unqualified to evaluate the work in question (Baldwin, 2018).[10-11] More provocatively, scholars of peer review's history warn us that, "The more we have expected of peer review, the more its opportunities to disappoint have expanded (Baldwin, 2018)." Our study here explores how widening the lens back from this historical contingency to include informal review might inspire more successful infrastructures for scholarly review.

## 3.2 Scholarly Skywriting: A brief history of informal review

Informal peer review, unlike its counterpart, has not been commented upon as much by historians, philosophers, and sociologists. Books, voluminous reports, and recurring conferences for its study are absent, though some reviewers have organized small, annual gatherings for an exclusive audience (Ramalho, 2024).

---

[9] Swedish humanities journals were late to embrace formal peer review, but evolved in the 1990s and early 2000s to require it (Müller et al., 2025). These journals had already practiced "proto peer review" wherein editors and externals referees informally reviewed manuscripts and acknowledged that scientists used more formal, codified procedures. Motived by local funders to standardize reviewing, a scholarly appreciation for impartiality, pragmatism, and strategic adaptation to elevate Swedish language journals in a landscape dominated by English-language journals, editors eventually adopted double-blinded formal peer review with external reviewers (Müller et al., 2025).

[10] We surmise that some naïve readers' confusions with the use of "peer" to describe those who evaluate research online originates from this historical shift to constrain the circle of credible experts from the broader populace to formally credentialed scholars within a subfield.

From what we've gleaned of its prehistory, its earliest incarnations date to top-down crowdsourced initiatives like the development of computer networking in America. The telnet system at the New Jersey Institute of Technology (NJIT) and its Electronic Information Exchange System (EIES) led to four pioneering digital publications by 1996 – a read-only newsletter for communications, the open journal *Paper Fair*, the computing journal *Mental Workload,* and the journal *Legitech* (Bingham, 1999). The latter three periodicals were imaginative experiments for their time, as they allowed *readers* to comment on publications online. In *Legitech*, readers' comments and questions were summarized into a structured document to inform scientific advisers to the government.

These early examples of what is now called open (public) peer review and post-publication peer review spawned *from journals*, a community that science reformers now regard as overly traditional, formal, and secretive. These efforts may be thought of as the prehistory of informal review, as they relied upon journals to solicit engagement instead of publisher-independent social networking platforms.

When the World Wide Web launched, innovations like the High-Energy Physics Eprint Server and the World Journal Association toyed with, respectively, abandoning peer review altogether and replacing it with mandatory article scoring by readers (Bingham, 1999). These early endeavors, for social and technical reasons, met mixed fates. World Journal Association's democratic scoring system didn't replace formal peer review and when the *Medical Journal of Australia* tested a hybrid peer review process composed of top-down and bottom-up elements from 1996-1997, readers reviewed just 7/56 (12.5%) of articles post-publication (Bingham, 1999). These reader-reviewers, though, contributed novel insights absent from formal peer review, worked to publish the article quickly, connected to authors, and peered into the reviewing process in unique ways. That this occurred without formal training, financial incentives, or a culture of heavy online discussions by skilled computer users is strikingly successful.

In this interstitial time between difficult-to-use telnet technology and polished social media platforms, scholars and journals were inventive and optimistic for their time. Lacking theories and firm empirical grounding, they created interactive discussion features on their web pages for editors, authors, and readers. The journals *Electronic Journal of Sociology (EJS),* which backtracked its reforms citing low participation, and the *Journal of Interactive Media in Education (JIME)* pioneered threaded commenting and voting (Bingham, 1999). Until that point, alternative peer review with its public, post-publication commentary had largely operated within journals' jurisdiction. After the turn of the millennium, dozens of general-purpose social media sites proliferated alongside a handful of academic ones. PubPeer, an early academic forum for elevating anonymous whistleblowers and post-publication peer reviewers, launched in 2012 and became a popular speakeasy for commenters across disciplinary divides. Soon after its creation, a journal editor praised PubPeer for its ability to help identify errors, correct the record, and ideate extensions to studies (Townsend, 2013). Even general-audience platforms like Twitter were regarded favorably as a "virtual department" of colleagues that enabled informal "pre review" of ideas before publication (Darling, 2013).

Avid social media users, especially in the life sciences, began to appreciate informal post-publication reviews for its power to correct publications (Faulkes, 2014) but considered the tone and anonymity of some comments to be obstacles to their wider use. These scientists likened social media to a mega-conference in which reviews reified one of the most important functions of the scientific process – quality control. This framing of informal peer review's function opposes the original functions of formal peer review – safeguarding grant funding decisions from politicians and alleviating editorial load (Merriman, 2021).

In PubPeer's infancy, editors entertained open access to letters and comments for each article and posted peer reviews alongside editorial comments (Winker, 2015). Authors, it was argued, ought to respond to post-publication peer reviews beyond the "letters model still rooted in the era of the printing press" and make necessary revisions if they were sensible (Winker, 2015). More radically, they wanted hiring and promotion committees to recognize and reward researchers who engage in meaningful post-publication peer review. Nearly a decade since then, the desire to consistently reward any kind of reviewing remains largely unsatisfied, though more life scientists are volunteering to review preprints publicly (Avissar-Whiting et al., 2024).

The most recently mapped vision of formal and informal peer review is multifactorial: reviews in digital communities could be more modular, incentivized, integrated with other academic practices, and organized by communal norms (Boston, 2024). Reformers and tool-builders are only now experimenting with incentives like financial payments and integrations into routine academic practices like virtual journal clubs (alphaXiv, 2025; ResearchHub, 2024). Yet, connections to deep theorizing about informal review practices that could guide these design decisions remains rare. Our study aims to connect empirics to theory formation and practices to design directions.

## 3.3 Am I crazy or is this wrong? Empirical studies of informal review

Contrary to the rosy predictions of visionaries and tool-builders, the empirical evidence about publisher-independent reviews for online settings appears modest. A content analysis of PLOS articles found that scholars commented on just 4-22% of articles across PLOS journals with declining participation since 2009 (Wakeling, 2020). Readers commented more often than authors and about half of the readers' comments were academic discussions or critiques. Pre-publication peer review suffers from a similar lack of interest. On the life sciences preprint server bioRxiv and the medical science preprint server medRxiv, just 7.3% of preprints from 2020 received at least one comment (Carneiro, 2023). Depending on one's judgment of the quality and relevance of typical study reports, this may be perceived as too few or just enough comments. Most often, comments were critical; they corrected claims or suggested changes to methods, analyses, and interpretations of insights.

When scholars discussed reports on the review platform PubPeer, life scientists often highlighted signs of fraud like image manipulation and misconduct, whereas social scientists most often reviewed methodological decisions and critically debated the paper in other ways (Ortega, 2022). These comments were rarely positive, though that may be due to the narrow scope of discussion platforms and the option to remain anonymous. Anonymous and named commenters on PubPeer and related platforms can be prolific "super commenters" with a predictable cadence.

In studies of crowdsourced, journal-led peer review garnered "relative success" by *Nature*, *Wellcome Research*, *PLOS ONE*, and the *Journal of Media Practice* (Skains, 2020). *Nature's* 2006 trial was unique, unifying top-down formal peer review orchestrated by editors with bottom-up informal review by users. Of the 1,369 papers in the trial, 68 (5%) passed an initial stage of formal peer review and then opted into a second stage of public peer review. Out of these permitted papers, just five (8%) received comments despite editorial efforts to solicit attention. These comments were not useful to editors' publication decisions, and they felt as though they were "pulling teeth" to gain comments (Skains, 2020). Soon after this failure, Nature abandoned its trial. Twelve years later, the editor-led open peer review platform PubMed Commons met a similar fate following tight restrictions on commenting that led to low participation (PubMed Commons Archives, 2018; Tattersall, 2015). Of its 28 million indexed articles, only 6,000 received any comments (McCook, 2018). Years later, the software startup Peer Review.io (peer-review.io) folded after low user engagement (Heard, 2024). Whether they are equipped with institutional resources or not, peer review platforms mighty and scrappy are more likely to fail than succeed.

In contrast, experiments at the journals *WOR*, *SQ*, *F1000Prime*, and *The Disrupted Journal of Media Practice* received higher participation using Hypothes.is to annotate manuscripts. Whether this derives from communal differences or not is worth testing. It is also worth asking what counts as a successful adoption of informal review.[12]

In-depth case studies of journal-independent informal reviews are sparse and typically focus on glaring examples of research fraud (Irawan et al., 2024; Yeo et al., 2017). In a brief weekly news report on social media topics, *Nature* reported on a 2014 paper critiqued by a geneticist on Twitter and later elaborated upon in a journal article (Woolston, 2015). A deeper and more exemplary case of informal review arose when bloggers and Twitter users discussed the supposed evidence of arsenic-based DNA skeptically and triggered the retraction of a paper with shoddy methods and hyped claims. As blog posts on Retraction Watch reveal, this is a familiar occurrence that could be even more common with publishers' cooperation. A chemist and research integrity scholar, summarizing the same arsenic life study, concluded that "the role of social media in peer review fills us with optimism and hope... (Jogalekar, 2015)"

---

[12] One might consider all manuscripts and publications to be deserving of commentaries. Alternatively, skeptics may view the firehose of article feeds to be wasteful research. We encourage you to reflect on this, as it carries implications for developing and scaling successful solutions.

## 3.4 Circles of credibility: Conceptual lenses on informal peer review

Given that explicit conceptualizations and theories of informal review are sparse, we sought a set of conceptual lenses from infrastructure studies and computer-supported cooperative work to make sense of our data. These conceptual lenses will help preview their role as sensitizing concepts in our later analyses and data-driven theorizing.

Practicing informal reviewing in such a public forum resembles a "glass laboratory" in which scholarly work is visible from the outside looking in (Edwards, 2019). This glass laboratory is infrastructure made of "robust networks of people, artifacts, and institutions. (Edwards, 2019)" Unlike other systems that focus on "generating, sharing, and maintaining specific knowledge about the human and natural worlds," the glass laboratory in which informal review operates attends to the *evaluation* of knowledge.

More specifically, the glass laboratory may also be thought of as an *evidence infrastructure* with "specific technological settings and conceptions concerning what pieces of knowledge qualify as evidence proper in a community of scholars and beyond" (Harvey, 2016). Such an infrastructure is dependent upon other, preexisting infrastructural layers: libraries, Internet communications, and academia. Here, we use *evidence infrastructure* to mean the system of (often self-organized) reviewers, communities, and technologies that evaluates and shares knowledge publicly. Informal review can be usefully understood as the operations of a *peer production community*, whose definition we adapt from Benkler, 2006: activities by volunteers who self-regulate modular tasks.

Peer production may or may not be oriented around developing open-access materials. When the goal is to develop such materials, we are observing a *commons-based peer production community*. Adapting an existing description, from Benkler, 2006, we define it as: collaborative, decentralized, and nonproprietary online activities by volunteers who self-design modular tasks to ingest, digest, and synthesize public knowledge for public benefits.

Crowd contributors have assembled successful open-source systems like Wikipedia, the Firefox web browser, Linux operating system, and GNU software suite. Commons-based peer production is the opposite of property-based (and market-based) firm production, in which firms use top-down delegation of tasks and centralized decision-making to produce private benefits. Within the construct commons-based peer production, we are particularly interested in *knowledge commons*, whose definition we adapt as (Hess & Ostrom, 2006): shared knowledge resources between community members.[13]

As our three cases of informal peer review in the Background above showed, data do not lie inert after publication. Instead, they continue their *data journeys* as they are scrutinized by experts across subfields and disciplines (Leonelli, 2020). There is truly nothing akin to a final form for data like scholarly reports; reviewers, authors, and publishers coordinate corrections, expressions of concern, and retractions, morphing the data into other shapes.

As scholarly claims journey from journals and conference proceedings to digital communities like social media sites, they battle *data friction*, or "the costs in time, energy, and attention required simply to collect, check, store, move, receive, and access data (Edwards, 2010)."

Journals may be paywalled. Materials, data, and code may need to be requested multiple times using different technologies to better evaluate claims. Or else, open artifacts may need to be downloaded, transformed, and inspected for comprehension in the face of subpar documentation. As a corrective measure, informal reviewers work deeply to learn the sociotechnical structures and behaviors that made the data what they are now. In other words, they engage in *infrastructural inversion*.

## 3.5 Everything is computer: Technologies to scale informal review

Several entrepreneurs and scholarly societies have attempted to design practices and platforms to support informal reviewing, though failures outnumber successes. Recent failures like Peer Review.io lie in a graveyard of doppelgängers; others like gotit.pub, Reviewer Commons, and Paperstars are attempting to attract users (Gotit Pub, 2025; Paperstars, 2025; Review Commons, 2025). Some users are hesitant to cross-post content on the most established and secure platforms like PubPeer, wondering if there is any value and how their intellectual property would be licensed (Fried, 2024). For the most influential and veteran informal reviewers, cross-posting on bespoke reviewing platforms and publishers' commenting spaces is unpalatable as doing so would give up privacy, intellectual property control, and their audience (Gelman, 2024).

[13] For the sake of brevity, we will not use "knowledge commons-based peer production communities."

Tool-builders in this space, then, must solve how to: (1) persuade influential members into participating and (2) engage outsiders across disciplines and cultures instead of subscribing to the mindset "Build it and they will come." One user-friendly platform, the preprint commentary platform alphaXiv, allows users to annotate preprints and directly converse with study authors (alphaXiv, 2025). Solutions like these respect the temporal and cognitive resource constraints of busy scholars, lend an air of casual conversation unrestricted by publishers' requirements, and invite a mélange of commenters. Drawing on these successes and inspiration from our digital ethnography, we will list specific and novel design ideas to support informal review even more.

# 4 Ethics

Our study used and generated public data from social media sites like Twitter, Bluesky, LinkedIn, Reddit, PubPeer, and personal websites. We aligned our study with international research conduct laws, terms of service on the platforms we used, and our institutional review board (AoIR IRE 3.0 Ethics Working Group, 2020). Given that informal reviewers engaged in typical activities that would have occurred without our ethnographic interactions and the level of risk was considered minimal, our university IRB granted us Exempt status (#2187204-1). This enabled us to implement our study plan without full review. See Supplementary Materials: S1 for approval documentation, more details on our ethical precautions, and reflections on our engagement with users[14].

We have been monitoring the cases and study members daily to evaluate the impacts of our study. In our report, we quote study members and name them as anonymization is not possible when quoting publicly available posts. We considered the potential for harm to each observed member of the online community and ultimately concluded that their already risky behaviors signaled that they understood the tradeoffs of going public.

[14] https://doi.org/10.6084/m9.figshare.32048553

# 5 Methodology

## 5.1 Research Approach

After reflecting on the literature, we concluded that informal peer review has been studied at a high level with *thin description* methods like statistics and surveys and at a low level with *thick description* methods like case studies. Between these extremes, we realized an opportunity to collect data at an intermediate level of granularity using a longitudinal digital ethnography[15]. This method, we decided, was low-risk, low-cost, and high-reward with the potential to forage across disciplines, communities, and platforms without losing valuable contextual details.

Our research approach, grounded in social constructivism, generated and analyzed Internet commenters' posts to understand the dynamics of informal reviewing qualitatively. We did not attend simply to individual posts, but the broader context of multiple posts and replies across time and platforms. We did this within an expansive, distributed infrastructure spanning bespoke and general-use social platforms, scholarly disciplines, analytic software, and strategies. Using ethnographic epistemology, an approach that captures study members' subjective experiences in their culture, we immersed ourselves in social settings and sought to gain an insider's (aka *emic*) perspective.

To do this, we used our existing technologies and connections, built upon 20 years of personal social media use across two data sourcers (JP, JC), and domain knowledge of key users. We approached this study with some prior knowledge of informal reviews from about ten months of consuming, cataloging, and discussing cases casually. Due to the uneven work distribution, this report uses "I" to specify work by the primary data sourcer (JP) and "we" to specify work JP performed with the help of my co-author (JC).

[15] Digital ethnographies are also referred to as online ethnographies, net ethnographies, web ethnographies, and virtual ethnographies. Here, we use the term "digital ethnography" for its familiarity.

## 5.2 Settings

Following IRB clearance with exempt status, we began formally sourcing and analyzing data for analysis during Fall 2024.

Around late 2022 to early 2023, the Academic Twitter migration[16] began to drain engagement on the popular microblogging platform and "virtual water cooler" for academic discussions. One analysis of 15,700 social scientists on Twitter reported that academics engaged less often when counting the number of active accounts, new tweets, replies, retweets, and quote tweets (Bisbee & Munger, 2024). This digital lethargy paralleled a leak of over a million Twitter users, peaking after the 2024 US Presidential election when one million users flocked to Bluesky. Many others migrated to Mastodon and LinkedIn, though measuring the digital diaspora is challenging because data linking users across platforms are not available.

Anecdotally, we've found that many of our own Twitter companions are increasingly easy to find on competing platforms, especially Bluesky and LinkedIn with bridging resources like Starter Packs and Feeds developing in spurts. Our timing was fortunate, as we were able to track the decline and growth of informal review communities across platforms as well as new platforms for preprint review. For a detailed description of our daily work, see S3.

As we will learn from Insights below, informal peer review has delayed impacts on publications. So, after writing our first draft during Winter 2024, we waited to update our cases with publishers' responses to informal peer reviewers twice. The first time, during Summer 2025, we updated cases and monitored conversations between reviewers and publishers. The second time, during Fall 2025, we updated cases again and checked for adjacent chatter that may be informative for characterizing informal peer review. Doing so allowed us to make statements about the durability of informal review communities and reconceptualize our theoretical implications. We ceased updates during early November 2025.[17]

[16] This is also known as the Twitter takeover and X-odus. Creative monikers for the migration abound.
[17] In early March 2026, we preprinted our final draft of this study report to solicit informal reviews and during April 2026, we submitted the revised manuscript to a preprint review service.

## 5.3 Dataset: Sources & Generation

### *5.3.1 Procedure*

We sourced informal review cases bottom-up from algorithmic suggestions in our social media feeds. It was generally difficult to source informal review cases top-down, as social media data are not structured for searching by category and currently available AI search engines fail at this task.[18] For instance, searching only for cases involving reviews of creativity issues or clarity concerns requires hacking together a data pipeline with other software. At times, we succeeded in *some* top-down crowdsourcing of informal review cases when asking informal review communities for cases that aligned with certain themes.

I developed a daily habit of reading the subreddit r/science, personal feeds on Twitter, Bluesky, LinkedIn, and popular academic blogs like Statistical Modeling, Causal Inference, and Social Science. I also considered secondary opportunities like comment sections on journal websites, preprint servers, and scholars' websites. These sites sufficed as gateways, generating a still-deepening pool of 100+ informal review cases between two data generators.[19] In this study, we analyzed what we considered to be the *most diverse* informal review cases across dimensions like salience, geography, discipline, community, and digital platform. Ultimately, we attended to 10 key cases, two of which are referenced twice in Results below, totaling 12 mentions of cases in Results below. In addition, we analyzed eight meta-commentaries on the dynamics of informal reviewing.

For curious readers, we overview our procedures and descriptions of our instruments in Fig. 1 below and elaborate on their details in S4 and S5. In it, we also discuss our use of convenience sampling, informed grounded theory, sensitizing concepts, logistics, our memo process, how we refined our codes, and study modifications.

---

[18] To rapidly source informal peer reviews, we recommend the software service Altmetrics to find social media posts by positive, neutral, and negative sentiment in Altmetrics Explorer. Altmetrics launched this feature during June 2025, though it requires more development to specify the *type* of negative comment. We are currently exploring the possibility of building tools for accurate semantic search over social media feeds to curate informal review cases about specific topics.

[19] For the curious, we share our rough, unanalyzed collection of 100+ informal review cases here: https://semble.so/profile/infotainment.bsky.social/collections/3mquxihlvda2c.

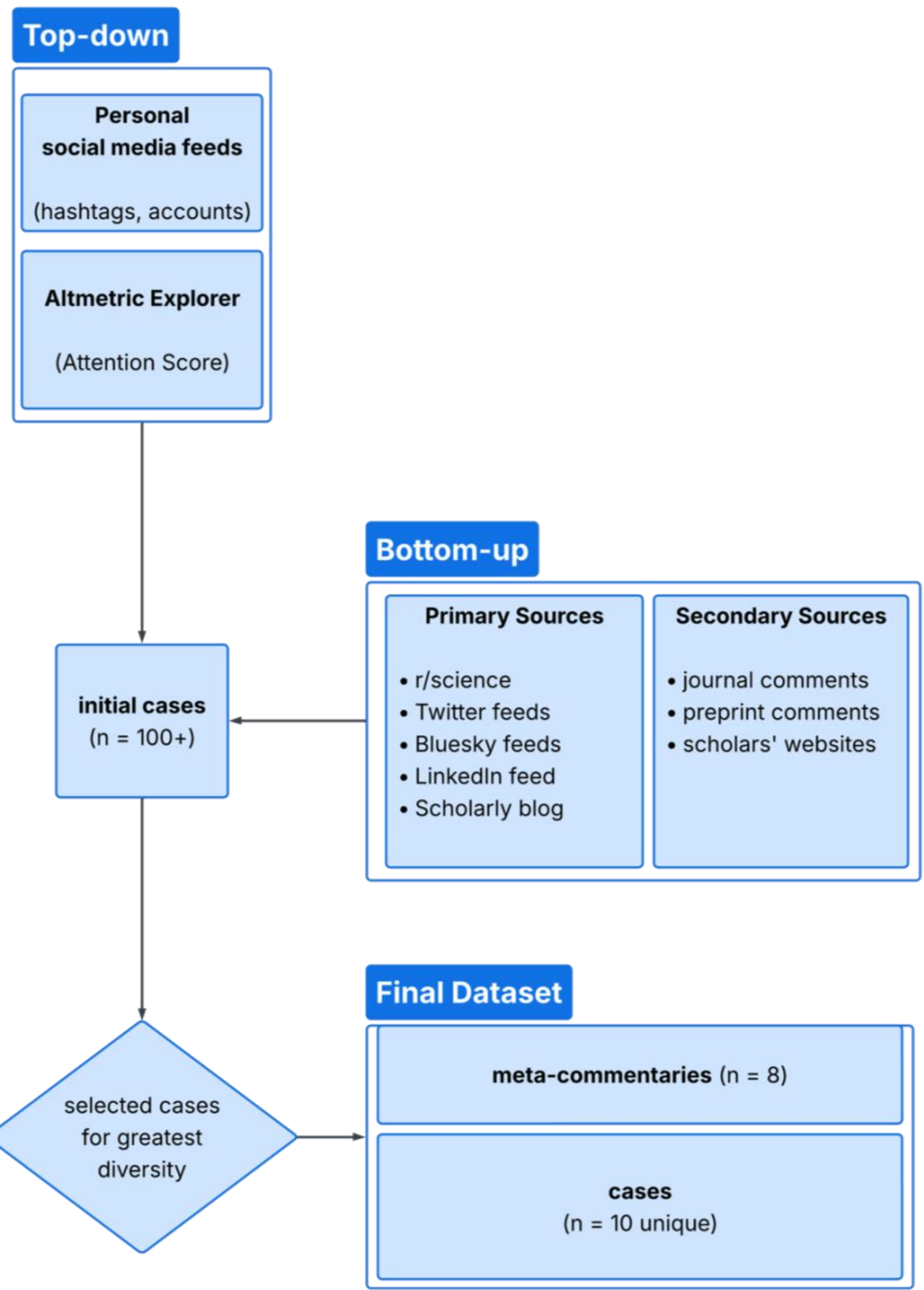

**Figure 1.** Data sourcing for informal review cases involved top-down sourcing by searching social media feeds and Altmetric Explorer and bottom-up methods like visiting social media threads and popular blogs daily. Filtering the resulting pool of 100+ cases for diversity of discipline, salience, geographic dispersion, and outcomes resulted in a final, analyzed dataset of eight meta-

commentaries and ten unique cases. We reported two cases, Metascience experiment and Cassava Sciences, twice in results. Hence, the total is 12 mentions of cases.

### *5.3.2 Active Participant Observation*

When I (JP) felt able to contribute, I asked questions, elaborated with praise or related resources, connected users within and between platforms, and facilitated community-building. Over the study period, I grew more comfortable liking, replying to, and sharing content from reviewers. These actions boosted the visibility of their content. At first, I lurked inconspicuously on blogs and around scholarly influencers on microblogging platforms. As I grew comfortable following and reading posts from informal reviews, I began discussing their reviews more deeply and shared my thoughts. These reviewers also began following my social media accounts spontaneously.

At this time, my profile on Twitter (@metasdl, now deactivated) and Bluesky (@infotainment.bsky.social, still active) included my name and personal details like hashtags on research interests, the current digital ethnography being conducted, and my institutional affiliations. On Reddit, HackerNews, PubPeer, and blog comments sections, I used randomly generated usernames or my first and last name without linking to my other social media profiles or institutional contact information. When possible, I transparently conveyed my identity as a researcher conducting a digital ethnography of informal peer review.

This dimension of the research was the most ethically sensitive, as I did not want to create or spread problematic posts. In each situation, I strived to interact with the scholarly community positively by minimizing unnecessarily conclusive statements and opening the discussion to alternative perspectives. I also tagged study authors, when possible, to allow them to clarify confusions and critiques. When considering potential errors that *I detected*, as in the case of a study with erroneous calculations of statistical power, I initially withheld posting an informal review and settled for an endorsement of my concern from established informal reviewers. Over time, I attempted to post to PubPeer and awaited my anonymous review to pass content moderation. Reading, questioning, and posting informal reviews in this way provided me with a deep understanding of this phenomenon.

### 5.3.3 Information Sufficiency

Finding diverse cases of informal reviews is trivial and the richness of this landscape invites a longer-than-feasible stay. In this study, sourcing cases until achieving full theoretical saturation was not practical as we were able to generate new themes and insights continuously. One would need multiple year-long ethnographies to describe the phenomenon fully. Instead, we segmented our data sourcing such that cases from July to November 2024 formed our first study report and the remaining, unanalyzed cases will be described in a second study report.

We sourced enough data to respond to our research questions substantively, connect to constructs in STS and metascience, and construct a theory of informal review that describes its character. This *information power* permitted us to crystallize four enlightening themes from a longer list (Malterud et al., 2016).

### 5.3.4 Final Dataset and Study Members

From our 100 candidate cases, we selected 32 that were detailed enough for discussion and potential analysis. Afterwards, we filtered these 32 informal review cases such that we focused on ten key cases, of which two are referenced twice in Results below (total: 12 mentions of cases). In all, we wrote fourteen memos about these cases. The textual data across these cases[20] totaled 125,472 words, including content from informal review cases found online. We also included eight meta-commentaries on informal review dynamics whose length reached 45,980 words. This dataset spans 21 focal study authors of papers being reviewed, sixteen distinct websites, and 27 commenters. Visit the open dataset on GitHub (https://github.com/oasisresearchlab/informal-peer-review) or Figshare (https://doi.org/10.6084/m9.figshare.33187710). These 27 commenters included informal reviewers and me (JP), as I conducted active participant observation.

Some of our study members call themselves sleuths and forensic metascientists, though we use the more neutral term *informal reviewer* in our study report to contrast with formal peer review and include commenters who fall outside sleuthing. For now, let's consider informal reviewers to be the broadest category and composed of anyone who critiques research online in a substantive way. When these reviewers intentionally seek out errors or issues multiple times and take on the identity of a careerist, they become a *sleuth*. Informal reviewers far outnumber sleuths, though it is necessary to conduct follow-up quantitative research to find out the precise discrepancy. For a detailed description of inclusion and exclusion criteria, see S2.

[20] Two of the ten cases were rich enough to mention twice in Results below; we reported on twelve instances of informal review cases (eight unique cases, two nonunique cases repeated twice).

I categorized the informal review data as text and images (data) from digital sources (settings). These 29 notes totaled 61,822 words, or about 4 hours of uninterrupted reading material at 250 words per minute. To understand the relationships and richness of cases, we organized all study data in an interactive node-link diagram (Fig. 2). In it, the circular *nodes* represent memo notes, and the straight lines are *edges* representing connections between memo notes.

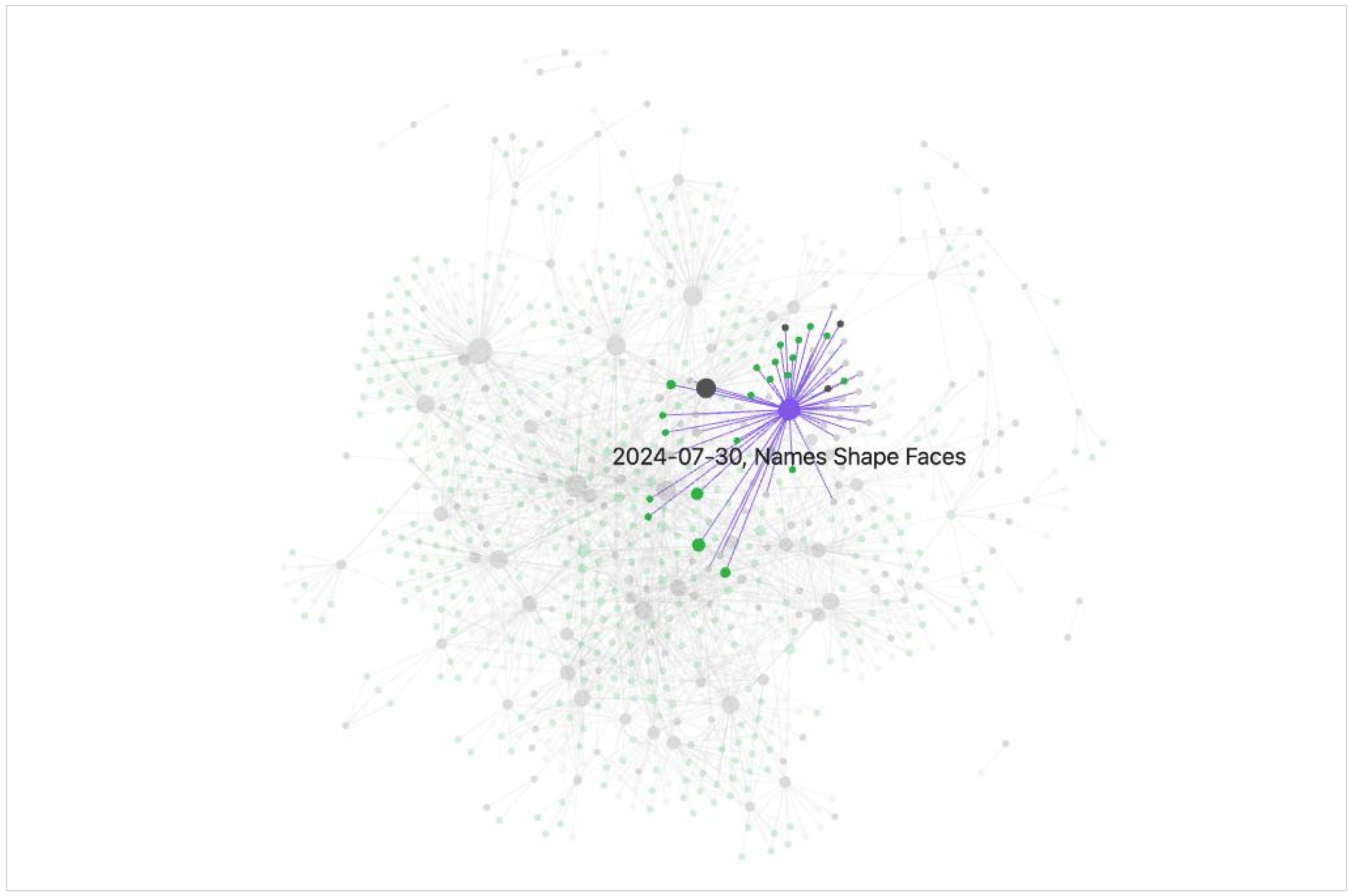

**Figure 2.** Interactive node-link diagram of the nodes (gray dots = case notes) and connections between them (lines = relationships between case notes expressed as hyperlinks in the notebook). The purple dots and lines are notes connected to a single case called *Names Shape Faces*. Other cases and their notes and grayed out and change color upon hovering and clicking.

### 5.3.5 Trustworthiness Practices

To build trust in our insights, we created audit trails whereby our raw data and analyses in the form of memos, codes, and a codebook. View the GitHub repository at https://github.com/oasisresearchlab/informal-peer-review or Figshare at https://doi.org/10.6084/m9.figshare.33187710. For a structured protocol that can be commented upon, forked, and shared, visit Protocols.io: protocols.io/view/protocol-informal-peer-review-patel-chan-2026-j6x6crfrf. We also asked study members and tool-builders in this area to comment on our themes by reading early versions of this study report. This private *member checking* process involved emailing and direct messaging users included in the cases described in the Insights below.

We have not triangulated our insights with multiple methods, though we are working within our research team to study informal reviews and the work using digital ethnographies, semi-structured interviews, and focus groups. These methods may provide supporting data across other informal review communities. It is also possible that network analysis and content analysis of large datasets of informal reviews will converge on similar themes that we describe in Results now.

## 6 Results

I (JP) used open and axial coding to generate a complex hierarchy of 1,080 codes that I refer to as the *coding tree*. At the first and highest level, I used the code #inpeer to refer to this informal peer review digital ethnography. Nested under that level in the second tier, I used the codes aligned with each of the four research questions, general observations, and notes about my engagement with the informal reviewers. For example, the code #inpeer/who refers to who engaged in informal review, #inpeer/evdInfra for evidence infrastructure issues, #inpeer/how to refer to how reviewing occurs, #inpeer/digitalEthnography to describe aspects of my participant observation, and #inpeer/why to organize content about why informal reviewers work. Each of these levels subsumes many other levels. I organized these codes and defined them in the open codebook with my co-author JC (see S6). Over two weeks, I worked with co-author JC to synthesize these codes into four themes of interest that were supported by multiple cases.

Each theme answers one of our four research questions, organized according to the context, process, and outcomes of informal reviewing. For each question-theme pair, we amassed a pool of relevant cases. To conserve space, we report on just three illustrative cases for each question-theme pair (12 cases covering 10 informal peer reviews overall).

## 6.1 Question 1: Who engages in informal peer reviewing?

First, let’s explore *informal reviewers* in three cases.

### *6.1.1 Case 1: Economic Review*

In this case, the historian and writer Rutger Bregman tweeted a paper titled “Reviewing studies of degrowth: Are claims matched by data, methods and policy analysis?” in the journal *Ecological Economics* (Savin & van den Bergh, 2024). That paper reviewed the economic literature of degrowth skeptically and found that few studies on this topic were empirical. Among the empirical studies, sample sizes were typically too low to make strong inferences. An academic expert on organization and sustainability, Jukka Rintamäki, quote-tweeted Bregman's post and replied to my (JP) probe to elaborate. He stated that the review did not use search terms broad enough to capture the corpus of papers on degrowth (Fig. 3). In a pithy review, he labeled the degrowth review as deserving of a “revise and resubmit.”

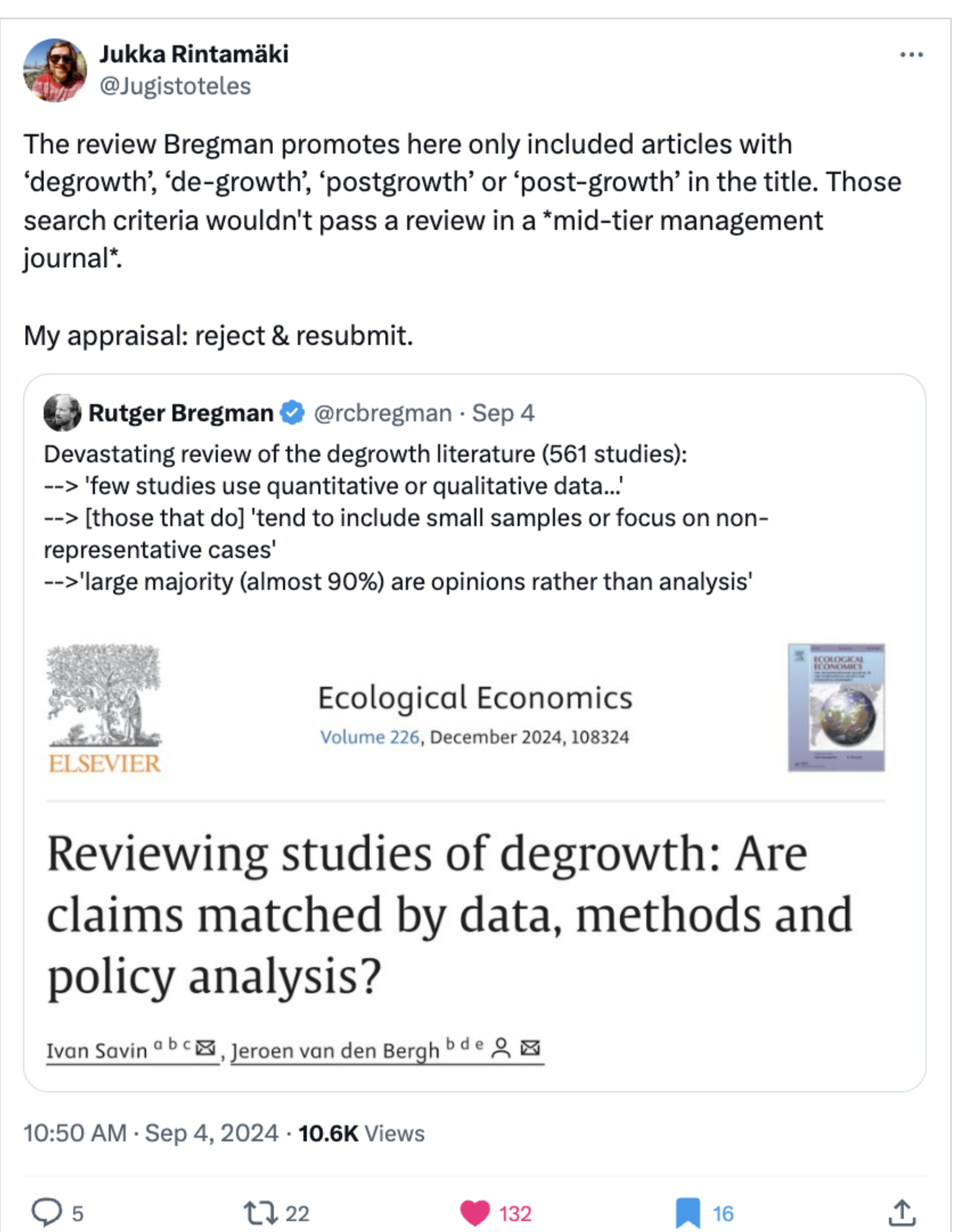

**Figure 3.** Informal review post on X/Twitter about a degrowth paper's narrow use of literature search terms (Permalink for original post: https://archive.ph/9joH5 and quote tweet: archive.ph/cHLc0).

In a separate quote tweet, an ex-physicist and current professor of climate and sustainability noted that the *Ecological Economics* review was too restrictive for ignoring computational modeling and qualitative papers (Fig. 4W). The study co-authors' epistemic gatekeeping undermined the foundations of their review.

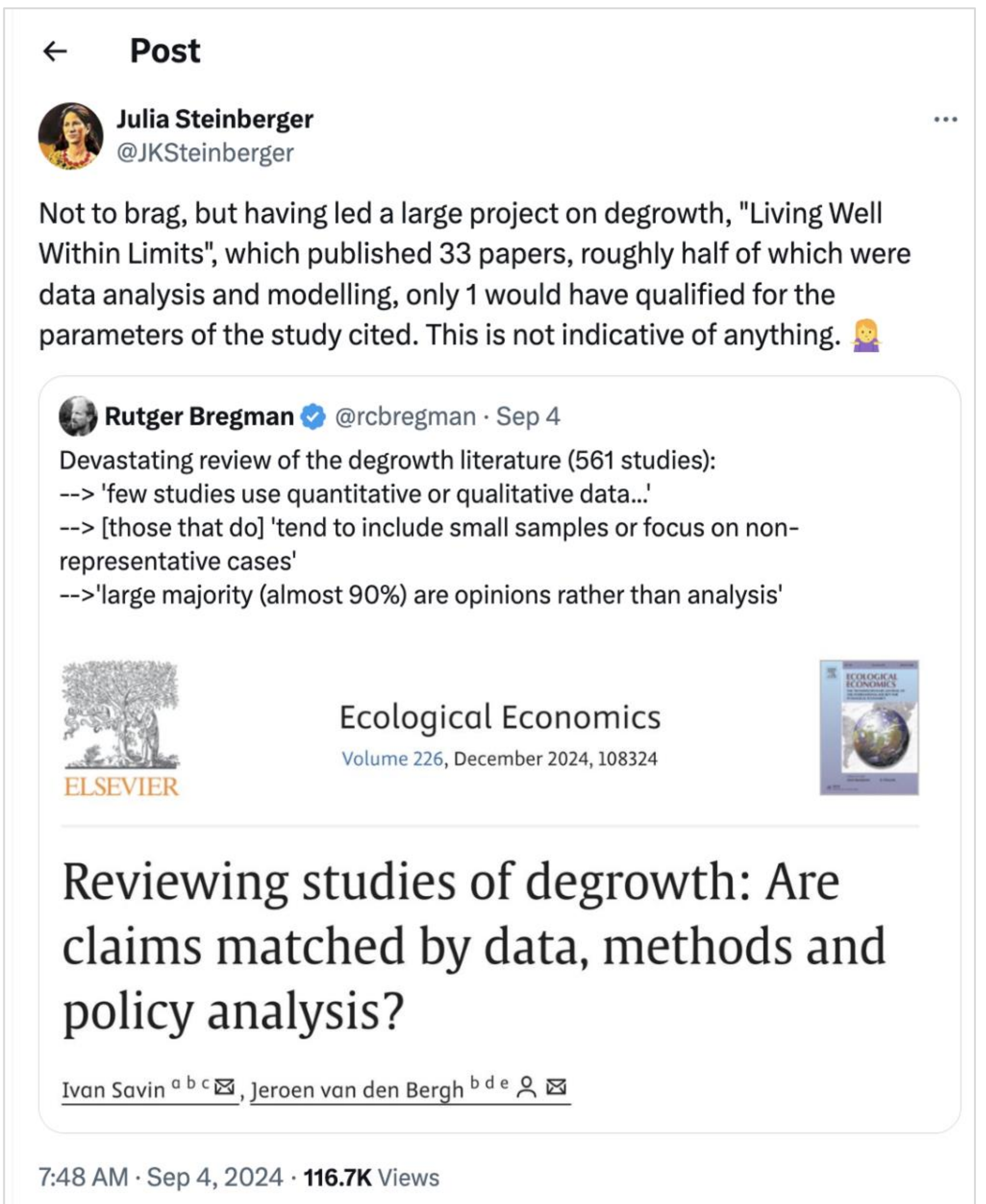

**Figure 4.** A climate scientist quote-tweeted a review on degrowth papers (Savin & van den Bergh, 2024) and concluded that the authors were too restrictive, neglecting many data analysis and modeling papers **(**Permalink: archive.ph/QvP2F).

Evidence synthesis expert and trainer Neal R Haddaway agreed that the review was poorly conducted (Fig. 5).

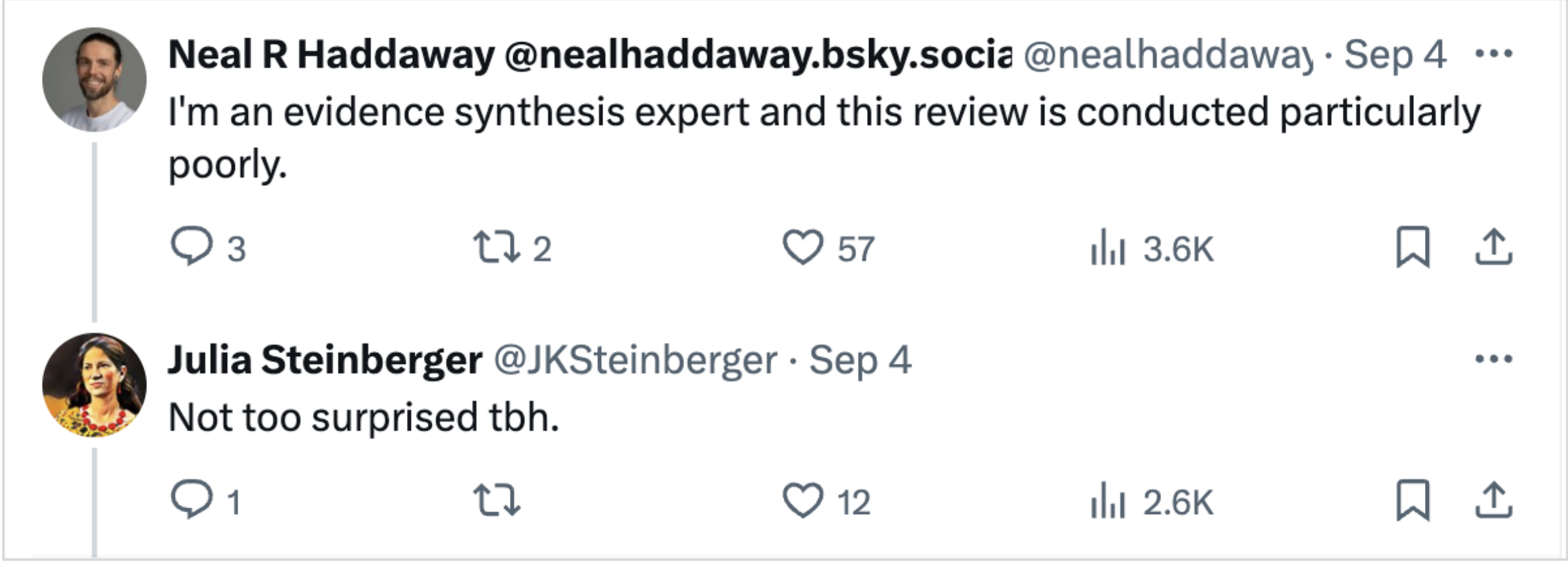

**Figure 5.** A brief informal review by an evidence synthesis researcher and trainer concludes that the degrowth paper reviewing the literature was narrowly scoped (Permalink: archive.ph/QvP2F).

Steinberger's thread went viral, being viewed 116,700 times, liked 1,100 times, and retweeted 250 times. This seemed to attract one of the co-authors of the review paper, Jeroen van den Berg, who replied to disagree with Steinberger by claiming that "a few additional ones [studies] wouldn't anyway alter the results."[21]

### *6.1.2 Case 2: Re-analysis of social media meta-analysis*

This longitudinal case, which crosses multiple communities and involves diverse stakeholders, was also viral on Twitter and Bluesky with news mentions in the academic periodical *The Conversation* (Meyerowitz-Katz & Jané, 2024).

The case begins with a meta-analysis, or quantitative synthesis of studies, of the impacts of social media use on adolescents' mental health (Ferguson, 2024). In it, the author concluded that the impacts of social media on adolescents' mental health were "nonsignificant." In response, psychologist Jonathan Haidt collaborated with his colleague Zach Rausch to re-analyze the meta-analytic data. After reading these reports, Ph.D. student Matthew Jané critiqued the lack of rigor in Haidt and Rausch's re-analysis using this statistical and psychometric skillset. He publicized his re-analysis of Haidt and Rausch's re-analysis of the Ferguson meta-analysis in a blog post, reporting that a more systematic re-analysis of the Ferguson (2024) meta-analysis showed that abstaining from social media has a null effect on self-reported teen mental health (Fig. 6).

[21] Twelve months after ending data collection, the paper in this case has not been corrected, updated, or otherwise revised on the webpage for *Ecological Economics*. This is expected, given the commonality of conducting cursory literature reviews. I (JP) engaged cross-posted the Twitter thread to the academic forum PubPeer (https://pubpeer.com/publications/DB2DA597904B3B3E8A74C0B0DB2DED#1) and summarized the informal review. The co-authors of the study did not reply to it.

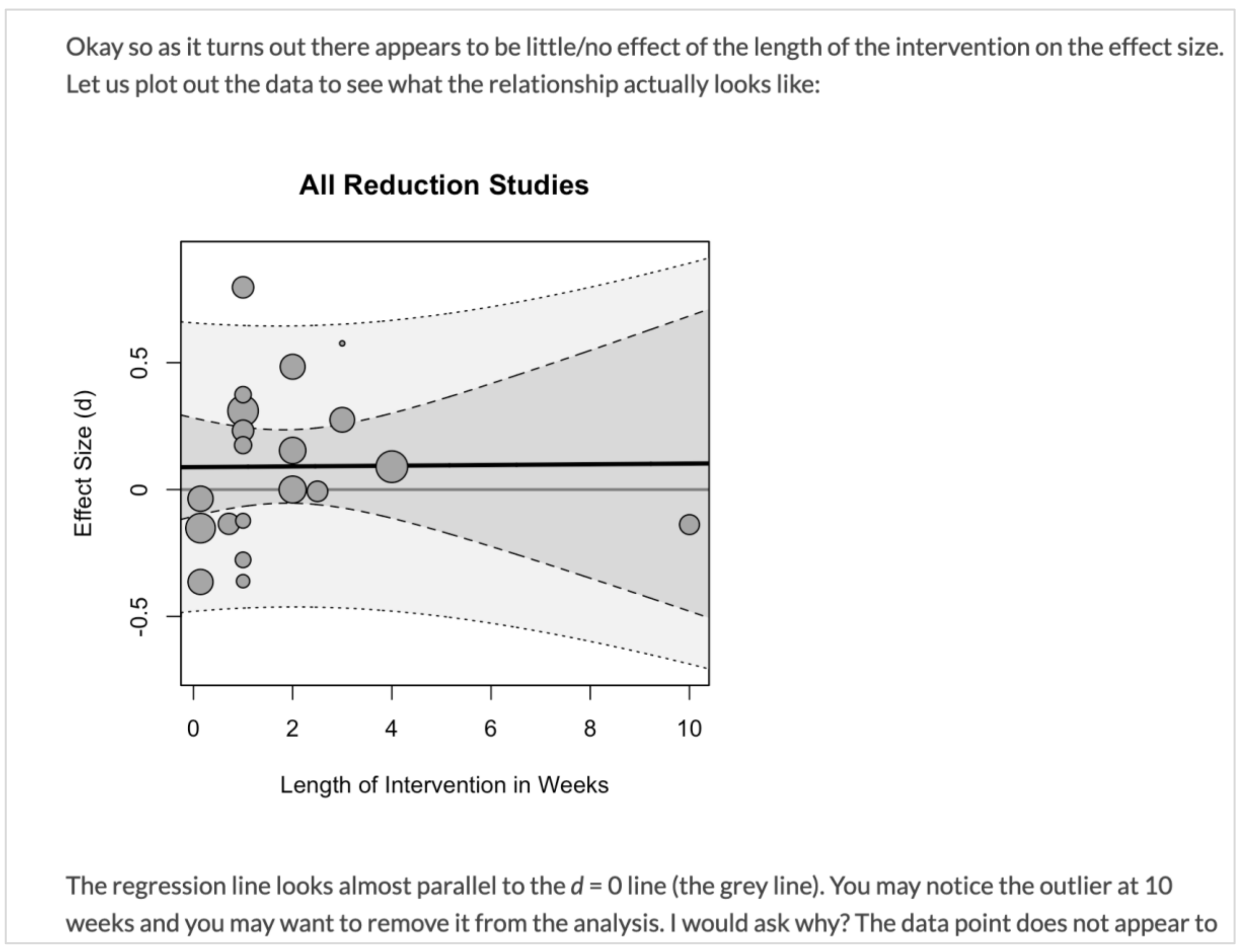

**Figure 6.** Excerpt from a blog post by Ph.D. student Matthew Jané, who re-analyzed the Ferguson meta-analysis about the effects of social media abstinence on mental health. He concluded that for interventions spanning 0-10 weeks, the effect of abstaining from social media on adolescent mental health is null (horizontal black line above). He has since published an extensive response to Haidt's response (Permalink: archive.ph/wCIYD).

In later interactions on Twitter and blog posts on their respective websites, Haidt and Jané conversed directly about analytic decisions (Fig. 7). Both parties admitted errors and limitations, ultimately producing an outline of a living meta-analysis that will be updated on his website as new data are available to synthesize.

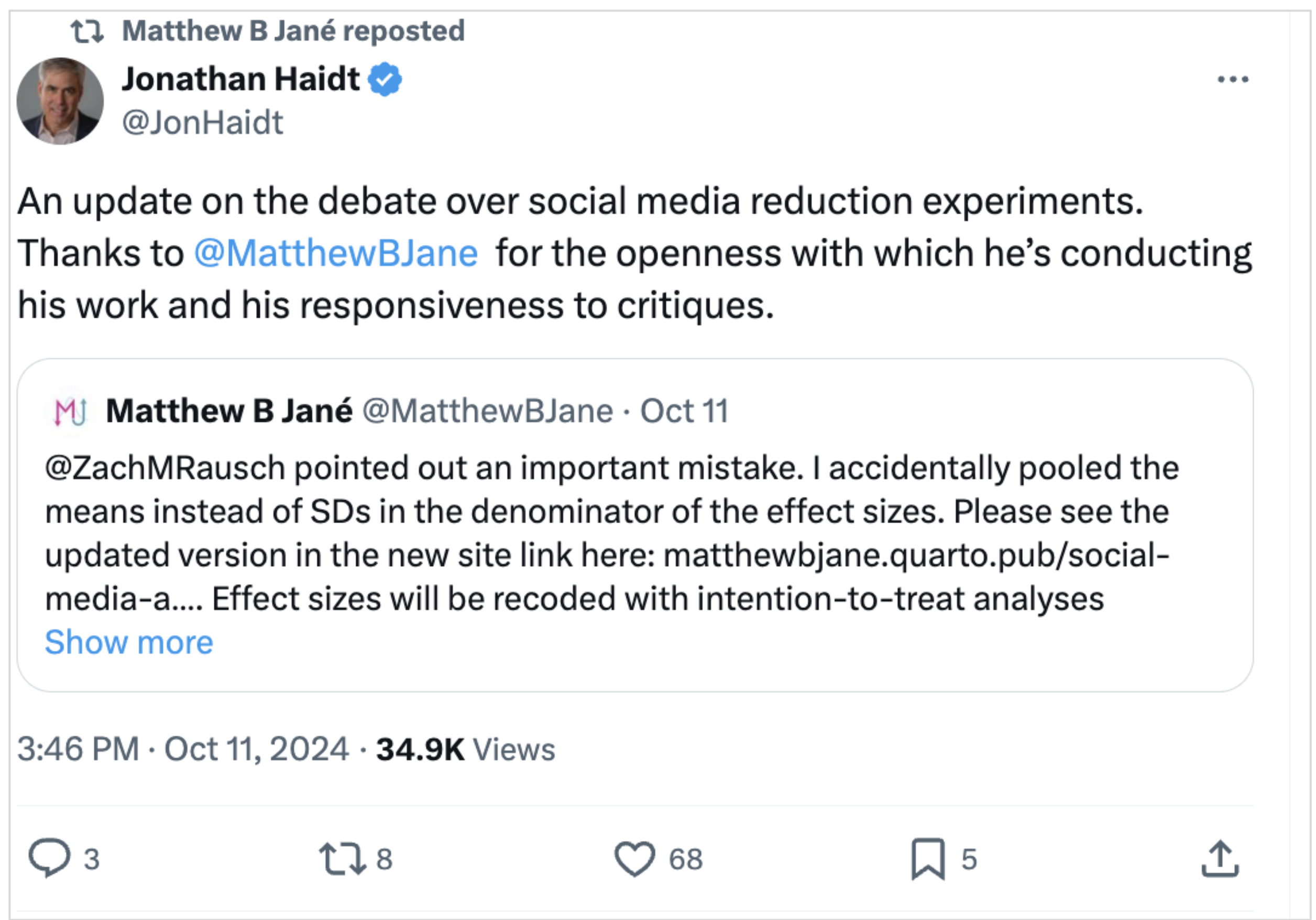

**Figure 7.** The Ph.D. student Matthew Jané, who re-analyzed a commentary by psychologist Jonathan Haidt admitted an error. Jané was able to update his living meta-analysis on social media's effects on mental health (Permalink: https://archive.ph/KMuFK).

Popular science communications like blogs and books pulled Jané's attention, not scholarly papers. In the periphery, professors in the social sciences scrutinized the strength of the average effect size and its meaning, rich discussion points yet to be published as formal reports.[22]

[22] Twelve months after ending data collection, Haidt's blog posts do not show acceptance of Matthew Jané's critique (Haidt, 2023). Jané, after earning accolades from the scholarly community for his re-analysis of social media and mental health meta-analyses, shifted to analyzing meta-analyses about public health issues like fluoride in drinking water to counter misinformation from political figures (https://pubpeer.org/publications/1A004FF18F5FA682CDC5D9899EB28F#2).

### 6.1.3 Case 3: Incorrect statistical power analysis

Problematic research also earns attention from retired academics working as volunteer sleuths to correct the scholarly corpus of erroneous, fraudulent, and doomed-to-fail studies. One sleuth, retired developmental psychologist Dorothy Bishop, often reviews studies with methodological and analytic errors on Twitter, Bluesky, and PubPeer. In this case, she reported that using incorrect defaults in a software program caused the authors to underestimate the statistical power of their neurostimulation experiment to treat obsessive-compulsive disorder (OCD). As she reviewed on PubPeer:

> The power of this study is overstated here: "The sample size was calculated a priori based on a medium effect size suggested for tDCS studies [42] ($f = 0.30$, $\alpha = 0.05$, power = 0.95, $N = 39$, mixed-model ANOVA with 3 measurements)." The final sample size was groups of 13, 12 and 12...
>
> Here's G*power output for this situation: (Note you have to specify SPSS in 'options' - see Lakens, D. (2013). Calculating and reporting effect sizes to facilitate cumulative science: A practical primer for t-tests and ANOVAs. Frontiers in Psychology, 4.

We've logged near-daily posts by Dorothy Bishop across informal reviewing communities focusing on methodological, statistical, and financial conflicts of interest in the psychological and neural sciences. She has the domain knowledge and statistical savvy to evaluate social-behavioral and neuroscience papers deeply.[23]

### 6.1.4 Theme 1: Motley crew

**Informal reviewers, it seems, are a motley crew of diverse scholars and hobbyists with wide-ranging employment status and affiliations across sectors**. Their skills and stakes in the research vary greatly to the extent that some of these reviewers would be passed over or dissuaded from formal peer review. They generate uniquely valuable

---

[23] Twelve months after a brief exchange on PubPeer in which the targeted authors deflected the critique, we did not find additional comments or revisions to the *Translational Psychiatry* paper titled "Targeting the prefrontal-supplementary motor network in obsessive-compulsive disorder with intensified electrical stimulation in two dosages: a randomized, controlled trial" (Alizadehgoradel et al., 2024).

insights that were absent during formal peer review, though scholars might question whether hobbyists, retired academics, scholars from unrelated disciplines, and others deserve to be called *peers.* We recognize this dilemma as an important one and return to it in the Discussion.

This motley crew extends across geographies and could be more diverse than the typical roster of formal peer reviewers. Elite credentials and membership in tight circles are irrelevant to participating across informal review communities. In some cases, a single case of informal review like a long, contentious thread on Twitter may attract an ensemble of agents – novices, specialists, study authors, and methodologists.

I (JP) did not find it surprising that informal reviews are so diverse and even expected this, though I was pleasantly surprised in Case 2 above that a young Ph.D. student blogging and even live-streaming reviews on YouTube earned virality as he debated an elite scholar (Jané, 2025).

## 6.2 Question 2: Where do informal peer reviewers work?

When we explored where the motley crew of informal reviewers works, we initially focused on individual digital communities like Twitter, Bluesky, PubPeer, and personal blogs. Later, we attended to informal reviewers' infrastructuring of specific subcommunities that we describe below.

### *6.2.1 Case 1: Bluesky Feed Scientific Comments*

Before I (JP) began sourcing data for this study formally, I stumbled upon an informal reviewer on the microblogging platform Bluesky who created the feed *Scientific Comments,* created by the tool *skyfeed* (skyfeed.app) and referenced by the hashtag #SciComment. The feed also picks up posts that include links to pubpeer.com and retractionwatch.com. On it, users are encouraged to discuss problematic research and scholarly corrections in a safe and civil setting. The creator, data visualization researcher Steve Haroz, enabled his community to connect in a way that Twitter did not permit and announced the creation of the feed to his followers in plain, terse language (Fig. 8).

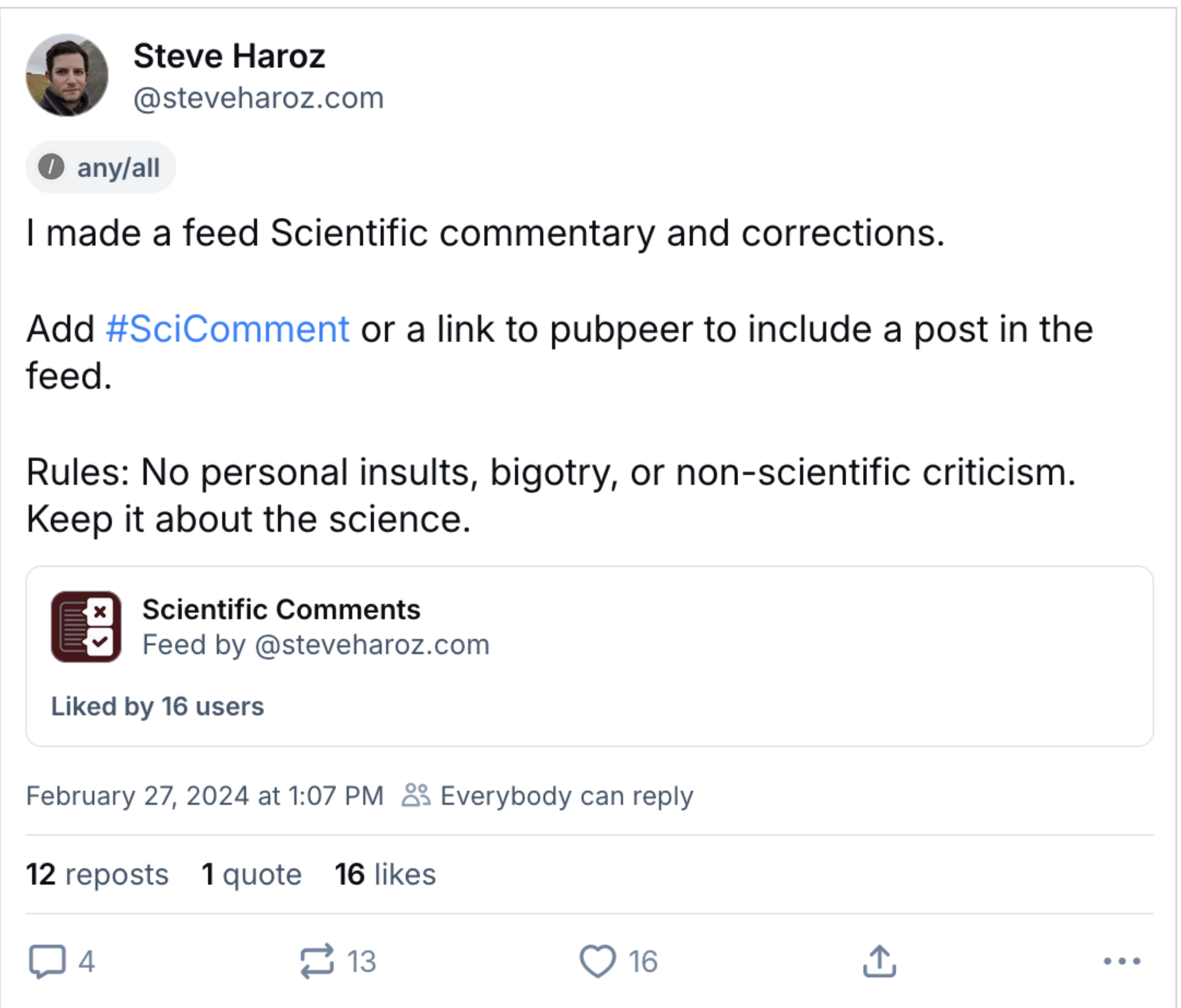

**Figure 8.** Inaugural Bluesky post announcing the creation of the Scientific Comments (#SciComment) feed for users to post relevant critical and civil posts (Permalink: https://archive.ph/wip/F3sNH).

Scientific Comments is a custom, user-created algorithmic feed. If Bluesky users link posts to the popular research integrity blog RetractionWatch or the academic forum PubPeer in their Bluesky posts, those posts will automatically crosspost to Scientific Comments (Fig. 9). In the year that I've spent browsing it, I've seen a geographically and disciplinarily diverse crew of users swarm to it and nest for long-term interactions.

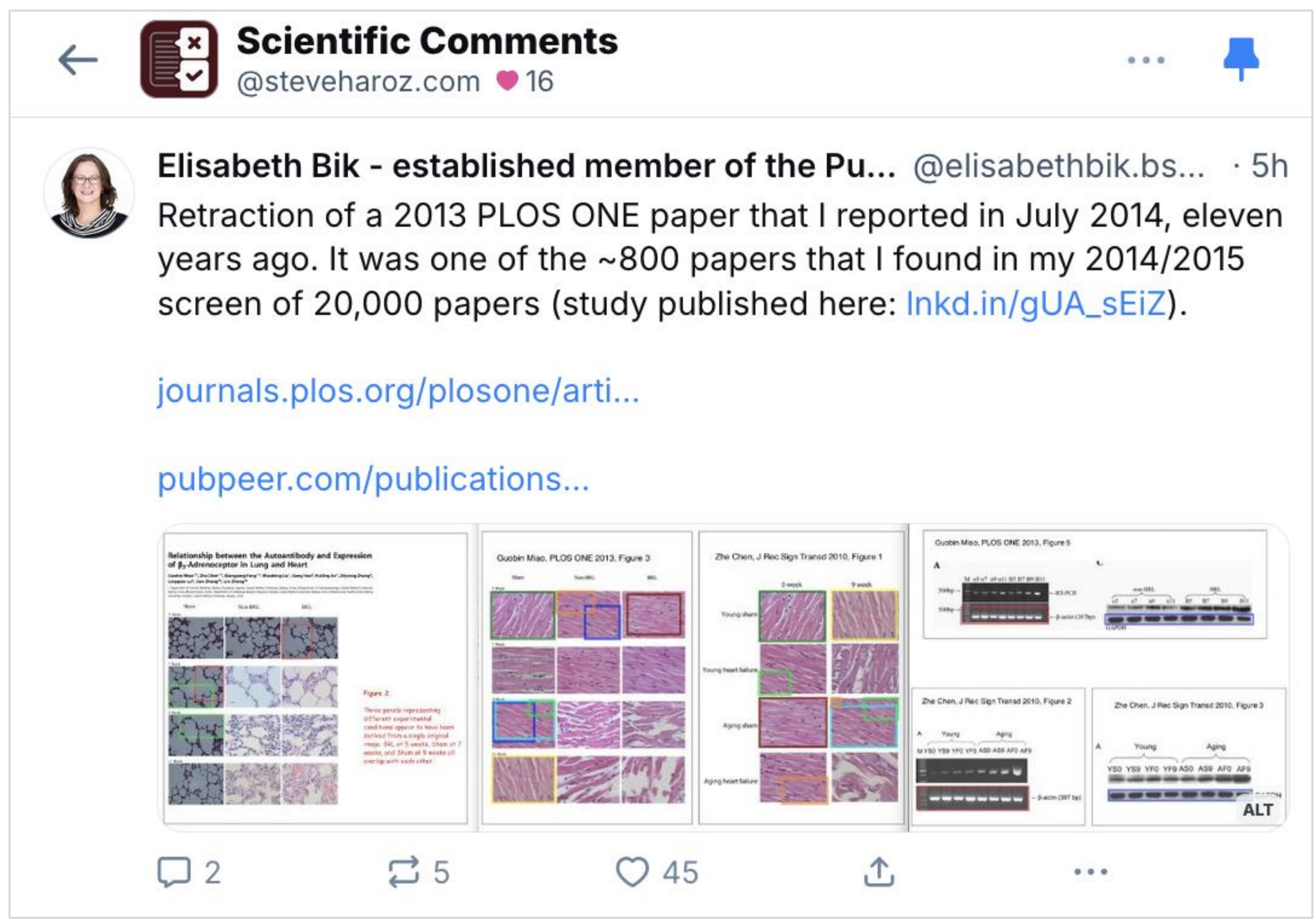

**Figure 9.** The Bluesky feed Scientific Comments includes the latest posts by diverse Bluesky users who share informal reviews and meta-comments about the practice (Permalink: archive.ph/Dxwdc).

Twelve months after being created, I observed that Scientific Comments still garners daily attention from a core base of followers and other users who post about research integrity, errors, questionable research, and metascience. Rising informal reviewers have entered the community smoothly, messaging veteran sleuths and contributing to statistical re-analyses of problematic publications in top journals (Hill, 2025).

### *6.2.2 Case 2: Bluesky Starter Pack #ResearchIntegrity*

During the aftermath of the 2024 US Presidential Election, the Academic Twitter migration dispersed users across competing social networks like Bluesky, Mastodon, and LinkedIn. To fertilize community-building on Bluesky, UK-based Research Integrity and Training Adviser Andrew Porter (@retropz.bsky.social) created the Starter Pack *Research Integrity* through which users interested in peer review, fraud, and plagiarism could network (Fig. 10).[24] Many scholars requested to be added to the Starter Pack and enthusiastically welcomed the community formation.

These Starter Packs are unlike Lists on Twitter. Whereas Lists are private collections of users, Starter Packs are public, capped at 150 users, and enable immediate induction into subcultures by clicking the self-explanatory "Follow All" button. As of our writing, Bluesky users have access to 324,516 Starter Packs and 108,151 Feeds to connect with personally relevant communities (Bluesky Directory, 2025).

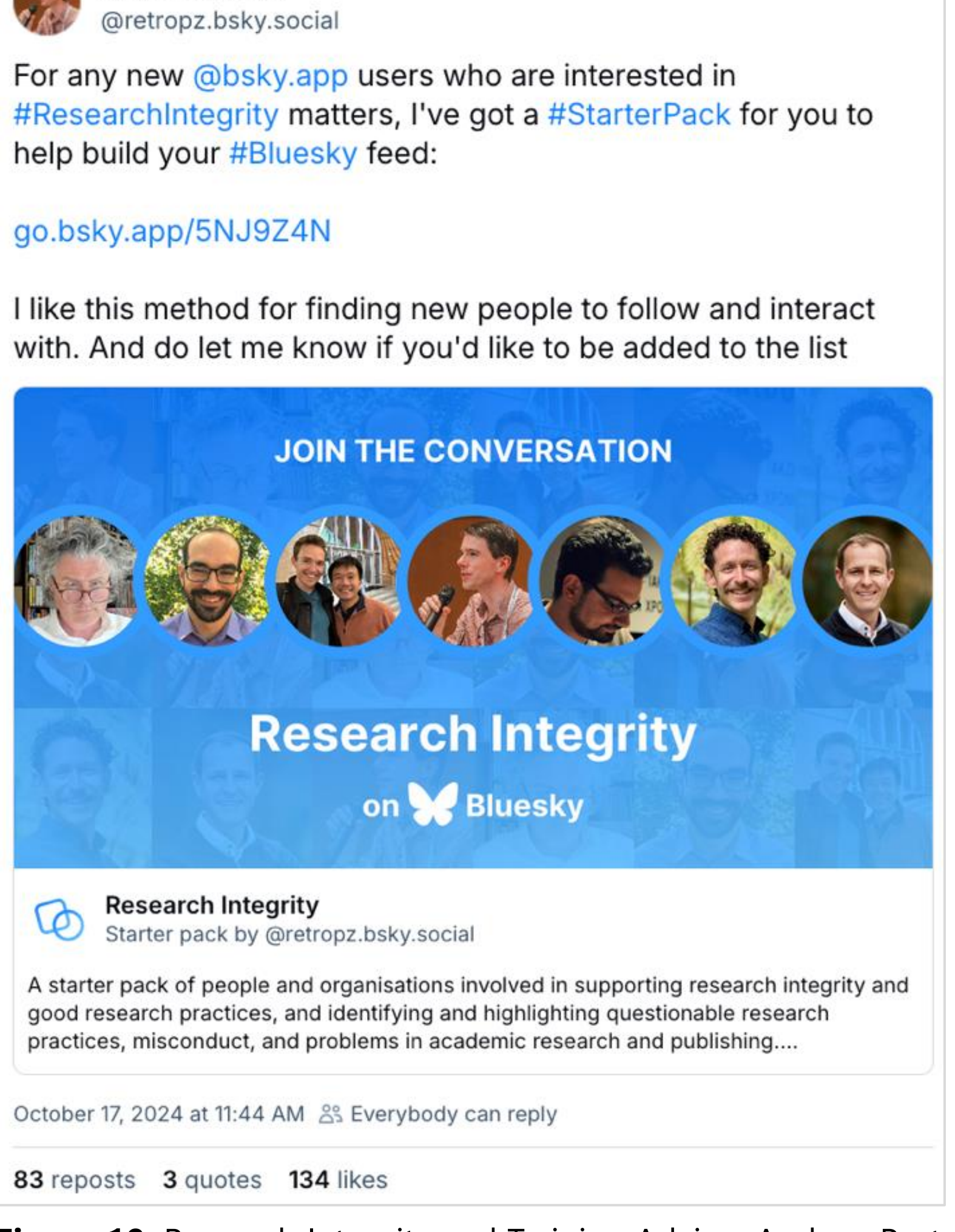

**Figure 10.** Research Integrity and Training Adviser Andrew Porter announced the creation of the #ResearchIntegrity Starter Pack on Bluesky to build community during the Academic Twitter migration in 2024. (Permalink: archive.ph/OExkh).

[24] Porter is based at the Cancer Research UK Manchester Institute at the University of Manchester.

As Porter and I both discovered, the open Authenticated Transfer Protocol (ATProto) that Bluesky uses enables Starter Packs to stream posts from Bluesky feeds. Early in the history of the Academic Twitter migration, informal reviewer Steve Haroz suggested infusing posts from Scientific Comments into the Starter Pack Research Integrity. Bridging technologies like these helped connect the huddled masses of the academic diaspora into a solid crew.

By leveraging available resources, resource-constrained volunteers interested in formal and informal reviewing practices *nest users* into relevant, discoverable subcultures and *bridge them* to common spaces like #SciFeed. The first Starter Pack was popular enough to fill the 150-member cap and resulted in a twin called #Research Integrity Part Deux[25]. About a year after being created, these Starter Packs are still active. The #Research Integrity Starter Pack and #Research Integrity Part Deux still bridge the feeds Scientific Comments and #ResearchIntegrity.

### *6.2.3 Case 3: Subreddit r/PeerReview*

Serendipitous searching one afternoon unearthed a fledgling *informal review community* on Reddit, r/PeerReview (reddit.com/r/PeerReview), from the influential sleuth, forensic metascientist, ex-academic, and startup founder James Heathers. As he announced in his Welcome to r/PeerReview post, he developed the subreddit and enlisted as its moderator because:

> It's become clearer in recent years that a lot of the best and most important peer review happens elsewhere - on PubPeer, in blogs, in newspaper columns, even on (God forbid) Twitter... great comments left in r/science... should have their own threads and discussions somewhere. So why not peer review on Reddit as well?

As a newer subreddit, r/PeerReview is composed of just 65 members including myself (JP). Its posts are by James Heathers and Redditors who informally review studies and solicit reviews[26]. These reviewers use their free time to discuss trending papers in brief and skeptical posts that attempt to balance civility with cheek. Just as the construction of this niche was informal and low-profile with an absence of posts on other social media sites, so too was its development. In his reply to my question about the launch, Heathers responded:

> I did not really announce it, no. The reason is: because it will take a long time to establish, one way or another.

[25] Research Integrity Part Deux Starter Pack: https://bsky.app/starter-pack-short/QN5upY5

[26] The subreddit r/PeerReview was created March 25th, 2025 and allows anyone to view and vote on posts. Submitting original posts is restricted unless approved by the moderator, who approves posts quickly.

For an overview of the subreddit r/PeerReview user interface, see Fig. 11 below.

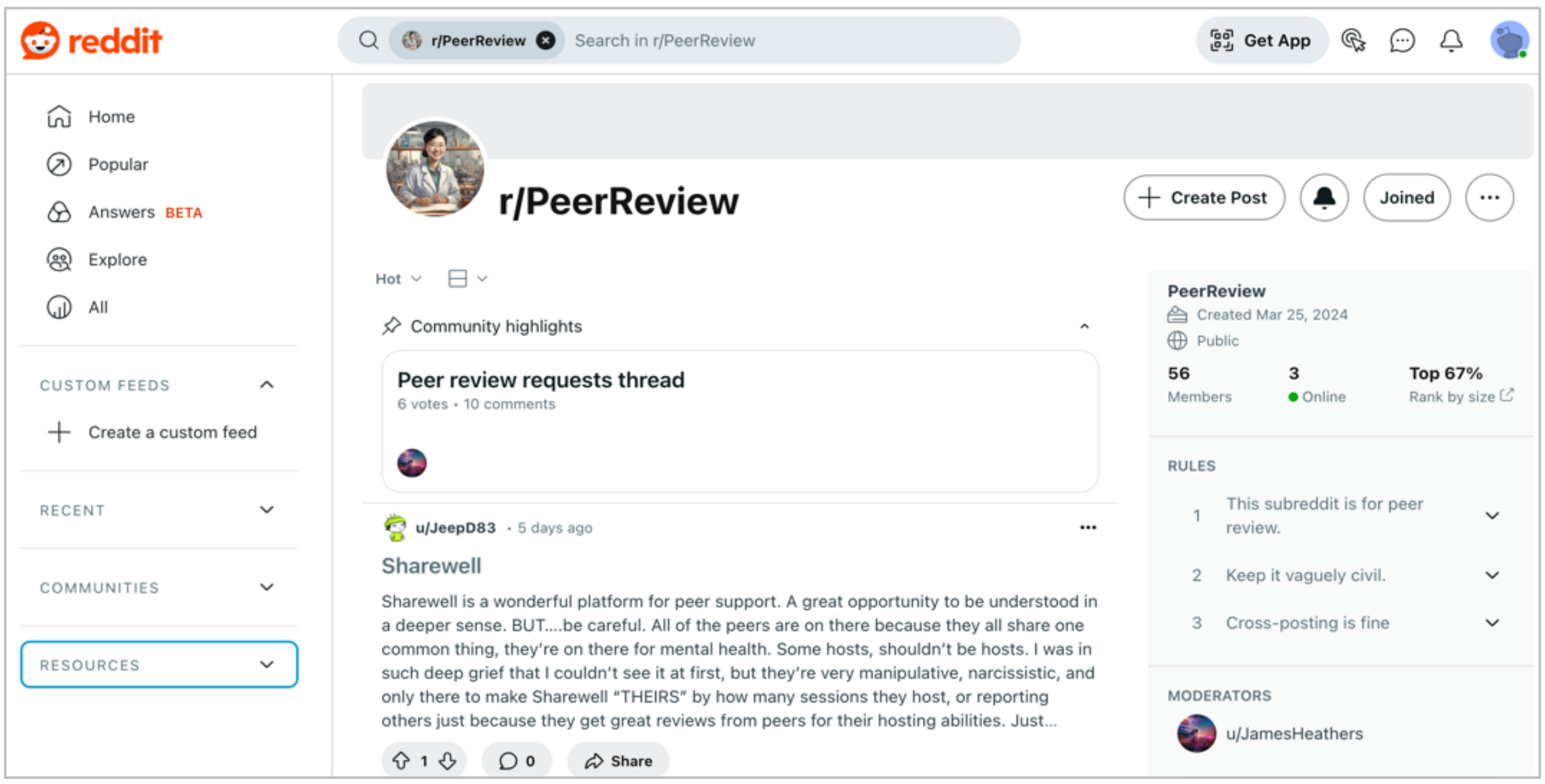

**Figure 11.** The nascent subreddit r/PeerReview, developed by the prominent informal reviewer, metascientist, and ex-academic James Heathers, has developed a small following (65 members) that reviews studies thought to be low-quality (Permalink: archive.ph/LlXtC).

Over a year after its stealth launch, there were few new and relevant informal review posts. Members authored just eighteen posts, excluding the inaugural post "Welcome to r/PeerReview." Some recent posts appear to be irrelevant and have garnered minimal engagement. In the meantime, Heathers has been busy launching other initiatives like a book on techniques for informal reviewers (whom he calls forensic reviewers) and a project to vet meta-analyses that are used to support medical guidelines (Heathers, 2025a, 2025c).

### 6.2.4 Theme 2: Hacking Homes by Nesting and Bridging

Across these three cases, we witnessed the motley crew of informal reviewers and supporters **develop their own infrastructure by hacking technologies and communities to develop low-cost, low-maintenance solutions with few volunteers, fragile technologies, nearly absent funding, and a shortage of institutional cachet**. In a classic paper defining infrastructure, we were advised to shift from asking *what* infrastructure is to *when* something becomes an infrastructure (Star & Ruhleder, 1996)

> The emergence of an infrastructure… is thus an "informal" one, evolving in response to the community evolution and adoption of infrastructure as natural, involving new forms and conventions that we cannot yet imagine (Star & Ruhleder, 1996).

As promised, the new forms and conventions of informal review infrastructure *were* unimaginable to us. Informal reviewers, taking the subtlest steps with humble tools, are hacking together their infrastructures with the knowledge that small wins may arrive, but virality and long-term security may be elusive even for the most influential reviewers. We are just now learning what is required to incentivize interested reviewers and the specific platform-level features that promise sustained, civil engagement.

## 6.3 Question 3: How do informal peer reviewers evaluate research?

Now, we explore if there are any patterns in *how* the motley crew of distributed informal reviewers evaluate research.

### 6.3.1 Case 1: Metascience Preregistration

In one of our longest, most publicized cases, an interdisciplinary scientist studying collective behavior and metascience, Joe Bak-Coleman, worked with other formal and informal reviewers to critique aspects of a high-profile metascientific experiment spanning years. This experiment was summarized in a paper titled *High replicability of newly discovered social-behavioural findings is achievable* (Protzko et al., 2024). The paper used modern, rigor-enhancing practices like pre-registration, open data, and high sample sizes to argue that most experiments in the socio-behavioral sciences *can* be replicated.

In a lengthy synthesis of his earlier formal and informal reviews of the *High replicability* paper, Bak-Coleman noted that he experienced friction accessing study materials. He needed to verify his critiques of the consistency of the preregistration with the executed study. He reflected on these difficulties in a blog post:

> Their OSF is such a mess that getting claims straight and conveying them to others is a nightmare. Many of the links above will just open a massive zip file you'll have to navigate to find the corresponding file... Other parts of their repo... are unlinked and can only be found by searching on the internet archive, finding a registration and working your way back up—often to a still private repository... preregistration seem to be little more than word documented uploaded to an OSF...the files are td;dr (too disorganized, didn't read).

Despite having access to study plans in the form of a pre-registration on the Open Science Repository (OSF, https://osf.io/), Bak-Coleman needed to exert hours of his time and considerable energy to locate the study planning documents, align them with claims in the publication, and author "an abbreviated version of 20+ pages [of] concerns I raised which were independently verified by a ten month investigation involving numerous domain and ethics experts."

Over twelve months after Bak-Coleman's blog post, the authors of the *High replicability* paper have not cleaned-up their open science files. Today, Bak-Coleman posts about the intersection of science, statistics, and society. He still remarks on problematic papers, though not with long-form analyses of the sort described in this case.

### 6.3.2 Case 2: Names Shape Faces

As with the case above, the Names Shapes Faces case[27] involved an informal reviewer with psychological, statistical, and metascientific expertise struggling to find a missing preregistration. In this case, multiple informal reviewers read the PNAS paper titled "Can names shape facial appearance?" skeptically (Zwebner et al., 2024). In the paper, informal peer reviewers noticed several red flags including a problematic editor, misinterpretations of statistics, and unclear statistical reporting (Daniël Lakens [@lakens], 2024). Here, we focus on a claim of having preregistered a key analysis for a claim mentioned in the study title. As Lakens remarked:

> Thanks! Yes, clearly an error on the journal side - it just highlights how much they care… The OSF link can not have this problem, it was just not made open - common, not too bad, it is just the overall pattern.

Less than a month after publication, perhaps due to prodding by informal reviewers, the journal updated the article with a correction and added the missing hyperlink to the OSF repository:

> August 9, 2024: The Materials and Methods section has been updated to correct a broken link.

As promised, the Materials and Methods section in Zwebner et al. 2024 reported where to access the data, analysis code, study instructions, and preregistration:

> Data files, analysis codes, study instructions, and the preregistration can be found on the Open Science Framework…

This minimal correction failed to respond to the major theoretical, methodological, and analytic problems voiced by the scholarly community across Twitter. Twelve months after I completed data sourcing, the paper remains on *PNAS* without an Expression of Editorial Concern, Correction, or Retraction.

---

[27] This case is the same as the second case in Background above.

### 6.3.3 Case 3: Cassava Sciences Fraud

As with Case 1: Metascience Preregistration above, Case 3: Cassava Sciences Fraud involved multiple informal reviews and resulted in career-damaging allegations. In the late 1990s, an industry-academic research collaboration led to the formation of a publicly traded company in the US (Cassava Sciences, Nasdaq: $SAVA). Since then, a bold set of Cassava Sciences studies claiming to discover the mechanism behind Alzheimer's disease and a therapeutic drug to treat it have been reported in a set of problematic peer-reviewed papers. In a conference poster presenting clinical trial results for the drug *simufilam*, the authors presented the following figures purporting to show a decline in harmful plasma P-tau proteins correlated with Alzheimer's Disease (Fig. 12).

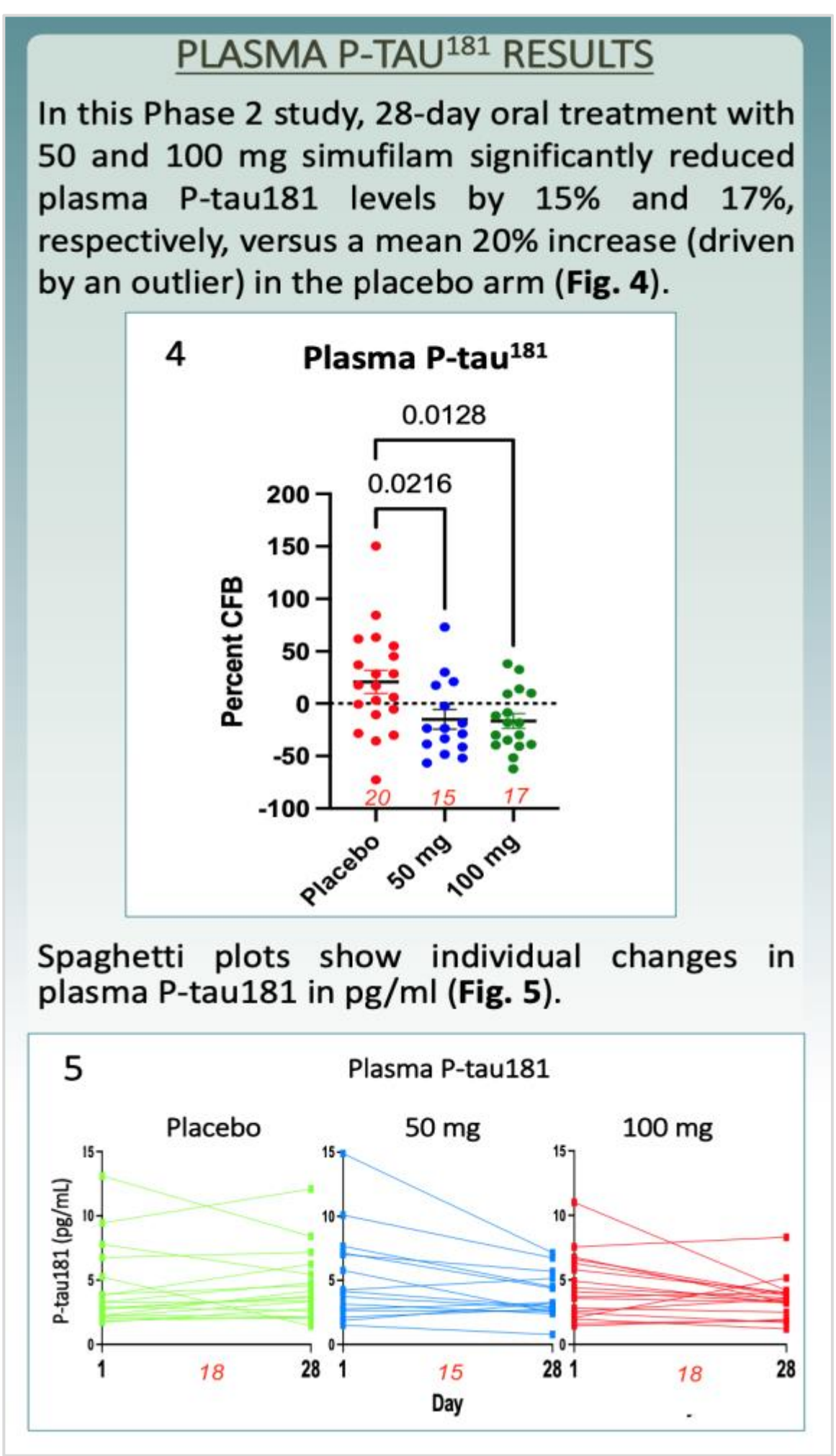

**Figure 12**. A problematic results section from a conference poster about the supposed benefits of using the Alzheimer's drug simufilam was targeted by several informal reviewers. (Permalink: https://archive.ph/wip/Lemp1).

Upon seeing this, the veteran sleuth and former microbiologist Elisabeth Bik used special analytic software (WebPlotDigitizer) to pinpoint the specific data points and contrasted them between figures (Fig. 13, colored text for clarity):

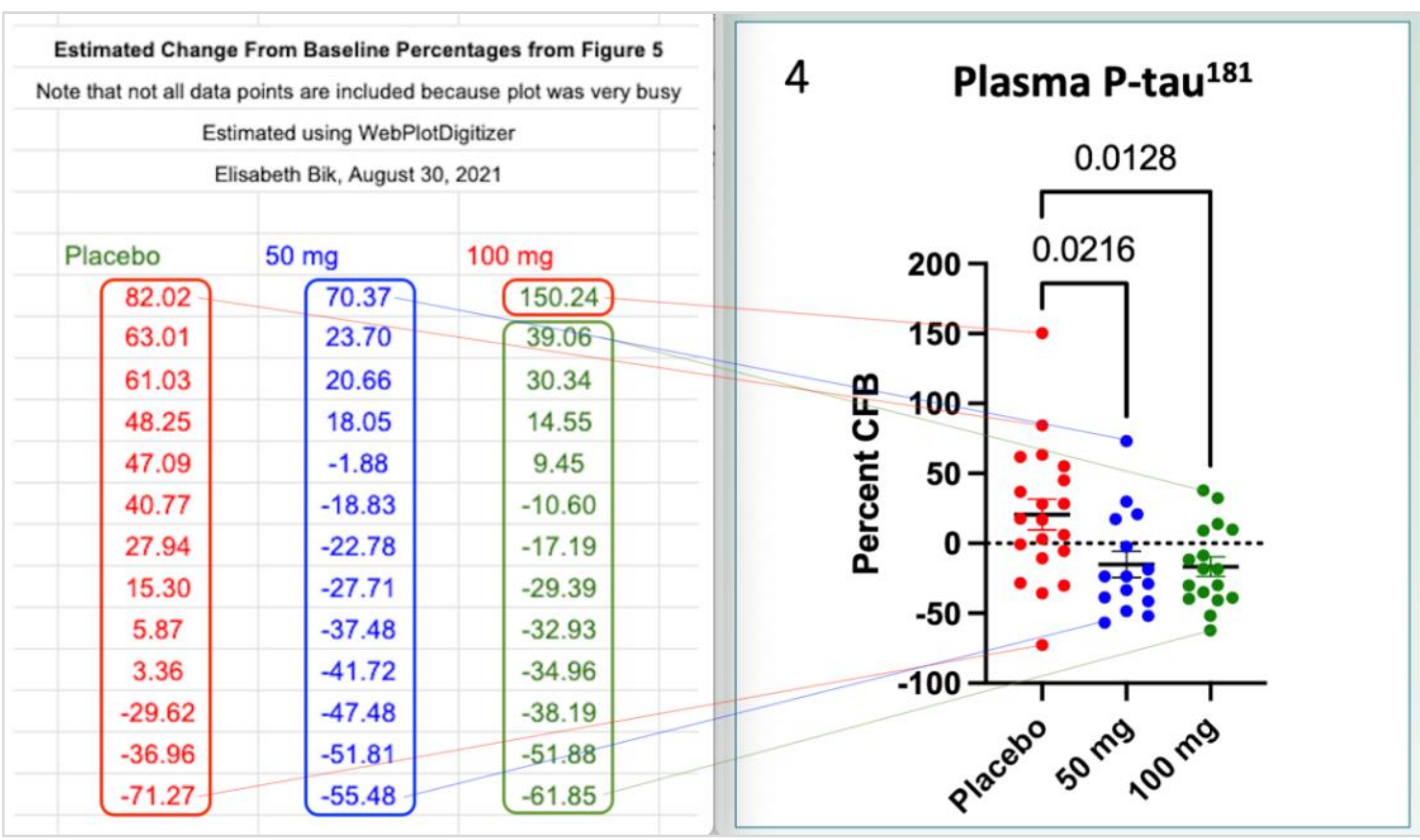

Estimated Change From Baseline Percentages from Figure 5

Note that not all data points are included because plot was very busy

Estimated using WebPlotDigitizer

Elisabeth Bik, August 30, 2021

| Placebo | 50 mg | 100 mg |
|---|---|---|
| 82.02 | 70.37 | 150.24 |
| 63.01 | 23.70 | 39.06 |
| 61.03 | 20.66 | 30.34 |
| 48.25 | 18.05 | 14.55 |
| 47.09 | -1.88 | 9.45 |
| 40.77 | -18.83 | -10.60 |
| 27.94 | -22.78 | -17.19 |
| 15.30 | -27.71 | -29.39 |
| 5.87 | -37.48 | -32.93 |
| 3.36 | -41.72 | -34.96 |
| -29.62 | -47.48 | -38.19 |
| -36.96 | -51.81 | -51.88 |
| -71.27 | -55.48 | -61.85 |

**Figure 13.** Cross-checking data from two figures in the Cassava Sciences poster on the supposed effects of the Alzheimer's drug simufilam indicated anomalous data points. On the left side, the list of numbers in Fig. 5 shows changes between placebo, 50 mg simufilam-treated, and 100 mg simufilam-treated lab rodents. On the right, a dot plot with connected color-coded lines indicates that in the 100 mg condition, the outlier point 150.24 in the red box was switched to the placebo condition in the dot plot (Permalink: archive.ph/wip/Lemp1).

> I did a more precise look at the estimated CFB percentages, by extracting the values in Figure 5 using WebPlotDigitizer... I focused on the lines that showed the biggest changes between day 1 and day 28, leaving out some lines that were horizontal-ish."
>
> Looking at the estimated change-percentages, Figure 5's estimated values appear to line up nicely with the position of the data points in Figure 4. Except, that is, for **the 150% data point**, which seems to have jumped from the **100 mg treatment group** to the **Placebo group**.

Here, the use of the software program WebPlotDigitizer was uniquely helpful in supporting deep analysis and ultimately discovering the presumed data falsification. This program scans points in data visualizations and extracts the values. In this case, the y-axis values for each dot can be extracted without requesting the official, original dataset. Downloading and using such a program isn't challenging, but it requires awareness of its features and the motivation to apply it.

### 6.3.4 Theme 3: Intensive infrastructural inversion

**Informal reviewing, as the cases above exemplify, can leverage technological tools like WebPlotDigitizer and cognitive-behavioral methods like checking preregistrations to uncover opaque data-generating processes.** Errors, missteps, and shortcuts manifest as reviewers continue digging, discussing with others, and confronting the triad of authors, publishers, and online observers. As one informal reviewer teaches us, to make discoveries and "…find the problems you have to look at the details in detail…" (Gelman, 2024a). Informal reviewers rub against frictions like missing links and messy databases. Their intensive activities are attempts to rewind the sequence of events from smooth and tidy reported results to missteps and malpractices in analyses, data sourcing, and even study planning. We call this *intensive infrastructural inversion,* following Edwards, 2010.

Metascientists and disciplinary scholars have decried the difficulty of recovering the data-generating processes of individual studies repeatedly in public and private (Minocher et al., 2021). In our small sample of informal review cases, we began to deeply understand the monumental difficulty of answering the question "How were these data and reports made?" Ostensibly, open science practices like sharing materials, data, code, protocols, and other supplementary artifacts would help, as would reporting guidelines that scaffold the transparent reporting of each step of the research process. In the cases below, we list the ways that even elites and science reformers guard against sleuths' attempts to trace the causal chain from claims and evidence back to their analytic and methodological actions.

From our early observations of informal review before conducting this study, we were aware that sleuths use a mixture of strategies and software programs. Still, we were surprised to observe the obstacle course of reviewing frictions and sleight-of-hand that they encountered when trying to access study materials and corresponding with publishers. Informal reviewers required months to years of sleuthing to notice and surface suspicious data. Inverting one's perception of the research infrastructure from polished papers and supporting materials to the people, tools, and technologies that engendered them is not always straightforward.

Sleuths are defined by the unique ability to invert the paper-producing scholarly infrastructure that pumps up weak claims and their dogged resolve to continue doing so despite social, temporal, and technological frictions. Much of their work is cognitive, often tedious. When reviewing tools *are* available, ones like statcheck ferret out statistical calculation errors and those like ImageTwin scrutinize microscopy images for signs of possible image manipulation (ImageTwin, 2025; Nuijten & Wicherts, 2023). Without them, errors and red flags would hide unbeknownst to the rushed reader.

## 6.4 Question 4: What are the impacts of informal peer reviewing?

Finally, we asked what good informal reviewing does using three sobering cases. Informal reviews have had a range of responses including ignoring issues, immunological self-preservation with evasive actions, minor corrections, major corrections like retractions, and retractions with a chain of cascading consequences.

### *6.4.1 Case 1: Apple Cider Vinegar*

A viral report titled “Apple cider vinegar for weight management in Lebanese adolescents and young adults with overweight and obesity: a randomised, double-blind, placebo-controlled study” was published in the *British Medical Journal: Nutrition (BMJ)* (Abou-Khalil et al., 2024). It quickly earned multiple informal reviews across scholarly and digital communities. One thread critiqued poor statistical analyses and insufficient analytic reporting for the bold and trending claim that consuming apple cider vinegar significantly reduces weight. In an example of deep review Professor of Human-Computer Interaction and Usable Safety Engineering André Calero Valdez re-analyzed the study results and simulated the missing 95% confidence intervals around point estimates (Fig. 14).

## Comparison of between group differences using CIs

```
pdata |>
  # calculate 95% ci for each measurement
  mutate(weight_ci = weightsd/sqrt(sample_size_per_group) * t_value) |>

  # generate plot
  ggplot() +
  aes(x = week, y = weightm, color = group, group = group) +
  geom_point(position=position_dodge(1.5)) +
  geom_errorbar(aes(ymin = weightm - weight_ci, ymax = weightm + weight_ci),
                width = 1,
                position=position_dodge(1.5)) +
  geom_line(position=position_dodge(1.5)) +
  labs(title = "Between-group comparisons using theoretical 95% confidence intervall",
       subtitle = "Assuming normal distribution of weight in each group",
       x = "week", y = "weight (kg)",
       caption = "Data copied from Table 2 of the paper.",
       color = "Experimental condition") +
  NULL
```

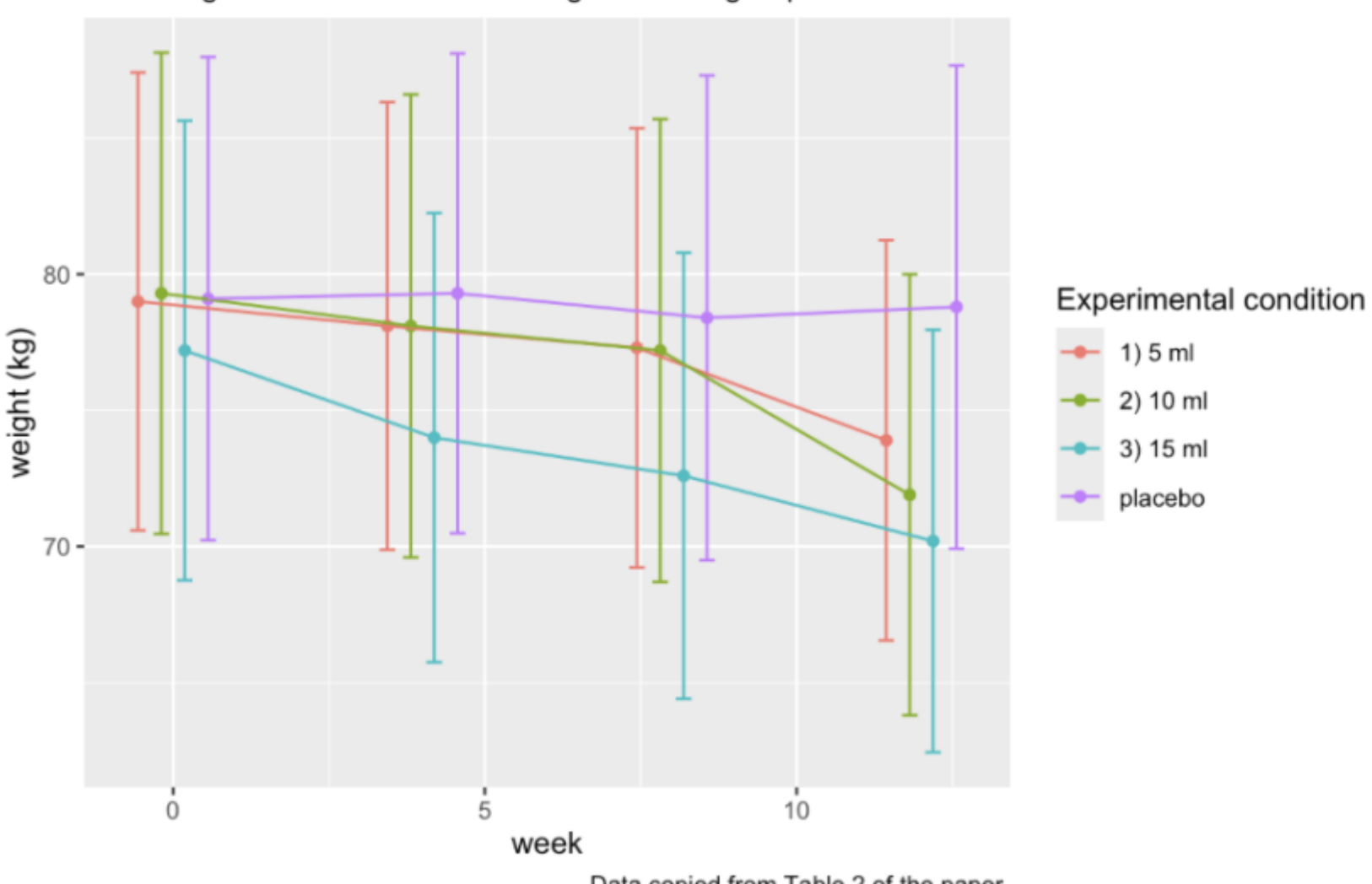

**Figure 14.** For the occasional informal reviewer Andre Valdez, the Abou-Khalil et al., 2024 paper claiming weight loss with apple cider vinegar defied belief. In a re-analysis of the data using means and standard deviations between four experimental conditions, he showed that the 95% confidence intervals overlapped. He concluded that over the treatment period, there were no significant intergroup differences in weight (kg).
Permalink: [archive.ph/submit/?url=https%3A%2F%2Fbmjnutritionacvdatanalaysis.netlify.app%2F](archive.ph/submit/?url=https%3A%2F%2Fbmjnutritionacvdatanalaysis.netlify.app%2F)

After posting on the journal's self-managed comments section called *Rapid Responses*, he claimed on Twitter and Bluesky that "Apparently, they have just ignored the rapid response, so far."

As part of active participant observation, I (JP), posted my own Rapid Response and emailed the journal to inquire about the delay in retracting the article. I learned from the response that the matter was currently being investigated by the Editor-in-Chief (personal communication). Over a year after publication and nine months after critical comments were posted in Rapid Responses, the study was still publicly available despite multiple persistent calls to retract it (Heathers, 2025b). Indeed, informal reviewers find that revising the scholarly record is typically difficult. They are occasionally treated to scholarly revisions, though most are hard-won after months to years of persistent, escalating messaging.

About 20 months after initial publication and eleven months after the first critical comment in the article webpage's Rapid Responses section, *BMJ: Nutrition, Prevention, & Health* retracted the paper for "the approach to statistical analysis of the dataset, implausible values, concern about the reliability of the underlying dataset, inadequate reporting of the work, and absence of prospective trial registration (BMJ Publishing Group Ltd., 2025). Using the provided dataset, the journal could not reproduce the results stated in the paper. The authors agreed to the retraction and *BMJ Nutrition, Prevention & Health* alerted the authors' institutions about the reasons for retraction.

### *6.4.2 Case 2: Metascience Preregistration*

After the informal review of the longitudinal metascientific experiment described in Question 3-Case 1, the corresponding author of Protzko et al., 2024 posted messages publicly reframing inaccurate claims as merely "an embarrassing error" in their communication (Fig. 15).

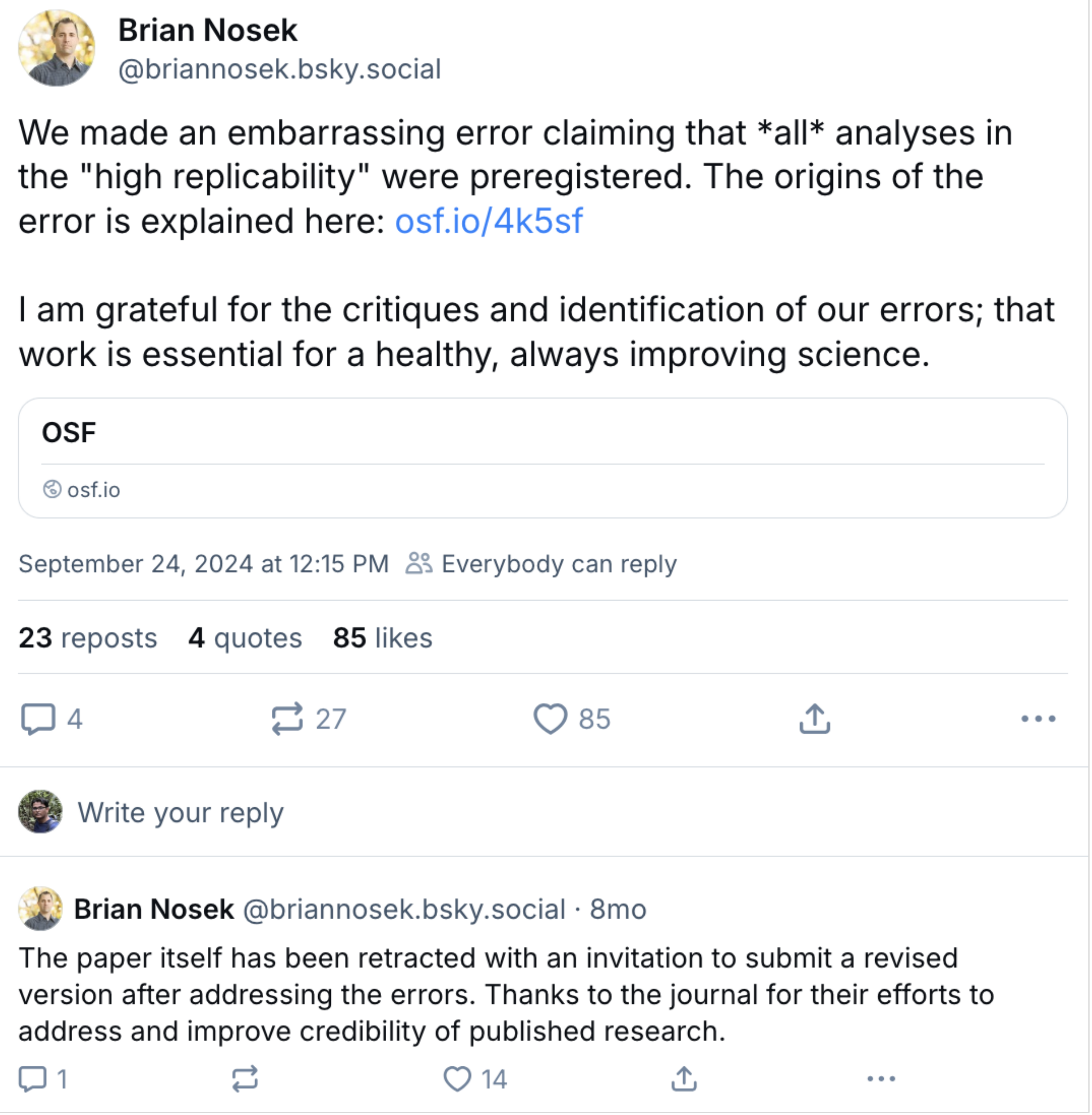

**Figure 15.** According to multiple formal and informal reviewers, a key study author of the retracted paper *High replicability of newly discovered social-behavioural findings is achievable* has been viewed as deflecting or downplaying his actions as errors. Permalink: https://archive.ph/oblJR

The journal that published the paper, *Nature Human Behavior*, opted for self-preservation when they allowed the authors to "resubmit a new manuscript for peer review." As they announced in the retraction notice to Protzko et al., 2024:

> Post-publication peer review and editorial examination of materials made available by the authors upheld these concerns. As a result, the Editors no longer have confidence in the reliability of the findings and conclusions reported in this article. The authors have been invited to submit a new manuscript for peer review (Protzko et al., 2024).

A close examination of informal reviewers like Joe Bak-Coleman, who also acted as a formal reviewer for the same paper, reveals the difficulty of countering influential scholars. *Nature Human Behavior*, as alluded to in Fig. 15 above, offered the study authors an opportunity to retract the paper and resubmit a new version. One might brand this rare opportunity a retract-and-resubmit. Given that the case is high-profile and complex, we encourage you to thoroughly study all the posts in our dataset[28], our codes, and a blog post summarizing the case by a prominent sleuth observing the unfolding of this lengthy and viral case (Gelman, 2024). Twelve months after ending data sourcing, the narrative of "an embarrassing error" remains uncorrected with scholarly news outlets mischaracterizing the retraction, counter to the informal reviewers' critiques, as an example of the difficulty of "doing good science" (Else, 2024).

### *6.4.3 Case 3: Cassava Sciences fraud*

For the set of informal reviews in RQ 3-Case 3 above, the biotech company Cassava Sciences responded to informal reviewers' allegations of fraud in a report to the US Food and Drug Administration (FDA) dismissively:

> As a science company, we champion facts that can be evaluated and verified," said Remi Barbier, President & CEO. "This helps people make informed choices. It is important for stakeholders to separate fact from fiction, which is why we wish to address allegations head-on.

Pharmaceutical businesses preserve their reputation against informal reviews by *denying problems* or risk lowering the value of their stock. After the fraudulent studies connected Cassava Sciences staff were publicized and their academic scientist was ousted and the ouster of an academic scientist collaborating with Cassava Sciences, their stock prices fell precipitously to $10.17 per share.

[28] Search the open dataset using the tag #inpeer/evdInfra/impacts/reactions/authors/refute/selfpreservation/downplaysRetraction

Nine months after initially documenting this case, I followed up to examine the wide-ranging impacts of informal reviews. In a rapid string of responses to the failed Phase III clinical trial of the Alzheimer's drug simufilam, Cassava ended its Alzheimer's Disease research, reduced its workforce by 33% (or 10 employees), hired new technical leadership, and shifted its focus to repurpose simufilam for treating epilepsy. At the beginning of FY 2025, its stock price bottomed out at $2.19 per share and Cassava's defamation lawsuit against plaintiffs was dismissed as "retaliation for their public criticism of the research" (Heilbut et al v. Cassava Sciences, Inc., No. 1, 2025).

Twelve months after initially documenting this case, I followed up one last time to check for updates. The key academic scientist in the case, Hoau-Yan Wang from City University of New York (CUNY) went before the US Court of Appeals. But after the jury was chosen, the US Department of Justice (DOJ) dropped the case because key documents about the case circulated between the US Office of Research Integrity, CUNY, and the National Institutes of Health. One document, a letter from CUNY, concluded that its high evidentiary standards for making accusations meant that Wang would not be found responsible for scientific misconduct (Travis, 2025). When this happened, ex-Cassava executives celebrated the exoneration, Professor Wang retired in July 2025, and informal reviewers expressed their disapproval online.

### *6.4.4 Theme 4: Immunologic Self-preservation*

**Taken together, the cases Metascience Preregistration, Cassava Sciences Fraud, and Apple cider Vinegar express the consequential and multifarious impacts of informal reviewing.** Informal reviewers may trigger scholarly revisions like retractions, but contend with stakeholders who deflect, deny, and delay until only the most committed informal reviewers remain. They conjure *non sequiturs* and gaslight observers routinely. Publishers, too, can be barriers to timely scholarly revisions and accountability. As one astute Twitter user remarked as he observed a separate case:

> My experience is that once a journal publishes a paper the editors develop an almost parental attachment to it and will defend it to the death. (Stephen [@Stephen50000766], 2024)

In the same vein, a prominent informal reviewer reflecting on his own informal reviews quipped that "With journals, it's all about the wedding, never about the marriage. (Gelman, 2024)."

Study authors who committed fraud are susceptible to criminal penalties like fines, investigations by their employer, a domino effect of publication revisions, and reputational losses. Considering the typically slow operations of the modern scholarly publishing ecosystem, it did not strike us as surprising that many scholarly revisions require perseverance. I observed swift retractions far less often. When they did occur, they surprised reviewers.[29] Authors, publishers, and institutions revise claims and evidence just enough to avoid appearing unresponsive. At times, their reactions resemble an immune system defending itself against hostile alien invaders.

These three cases portend problems with developing digital infrastructure for informal reviewing. Authors, publishers, and scientific businesses are incentivized to uphold their claims despite mounting suspicions and allegations. Given this, simply scaling up the practice of reviewing with bespoke technologies is insufficient to filter most research waste. Reviews on social media, PubPeer, and elsewhere must be coupled with the force to break through the web of incentives and standards that resist revisions.

# 7 Discussion

## 7.1 Summary

Originally, we asked ourselves four key questions about the structure, process, and outcomes: (RQ1) Who engages in informal peer reviewing? (RQ2) Where do informal reviewers work? (RQ3) How do they implement informal reviews? and (RQ4) What are the impacts of informal reviewing? To that end, we conducted a longitudinal digital ethnography to curate an initial pool of 100+ cases of informal reviews across Twitter, Bluesky, LinkedIn, PubPeer, ResearchHub, and blogs. In doing so, we elevated a phenomenon that has been studied only by content analyses and case studies.

[29] For a given problem or set of problems surfaced by informal peer reviewers, one wonders *whether and when* a scholarly revision will occur. We do not know for certain, but think that exploring this question in a follow-up study with content analyses of many more cases would be of immense value.

At this intermediate layer of description, our informed grounded theory methodology analyzing the twelve rich cases and eight meta-commentaries that we selected surfaced four enlightening themes worth reporting: (T1) Informal reviewers are a motley crew – highly diverse and inclusive of credentialed insiders and hobbyist outsiders, (T2) reviewing is distributed across public-access communities and platforms unused during formal peer review and clustered in niches using a mélange of free and simple resources, (T3) reviewers apply atypical technologies and reviewing strategies, and (T4) they persevere in their efforts to trigger unwelcome scholarly revisions. For a simplified overview of these themes, see Table 2 below and full details in S7, Table S1.

**Table 2.** Informal peer review cases and the themes that instantiate them.

| Themes | Cases |
|---|---|
| T1: motley crew | |
| informal reviewers are a diverse and distributed array of scholars and laypeople | • Degrowth Review<br>• Social Media Meta-Analysis<br>• Incorrect Power Analysis |
| T2: hacking homes | |
| informal reviewers often nest in niches like Scientific Comments and bridge platforms | • feed: Scientific Comments<br>• Starter Pack: #ResearchIntegrity<br>• r/PeerReview subreddit |
| T3: intensive infrastructural inversion | |
| informal reviewers evaluate research reports deeply, catching formal peer reviewers' errors | • Metascience Preregistration<br>• Names Shape Faces<br>• Cassava Sciences Fraud |
| T4: immunologic self-preservation | |
| study authors and publishers evade informal reviewers' critiques by denying and deflecting | • Metascience Preregistration<br>• Cassava Sciences Fraud<br>• Apple Cider Vinegar |

## 7.2 Triangulation and Interpretations

Perhaps because of the choice of research methods by scholars in this area, we have not read explicit statements aligned with all four themes: motley crew of reviewers, hacking homes by nesting and bridging communities, intensive infrastructural inversion, and immunologic self-preservation of authors and publishers. Existing empirical studies, which are content analyses of preprint comments and PubPeer posts, have reported on the frequencies of reviewing activities (Carneiro et al., 2023; Ortega, 2022). Case studies and sentiment analyses of problematic papers in the natural sciences are stimulating reads and display the power of what its early chroniclers called "informal post-publication peer review" and "social media-based, citizen-enabled peer review." These cases, though, have not treated us to an aerial view from which the infrastructure is discernible *as a whole* (Jogalekar, 2015; Meskus et al., 2018; Yeo et al., 2017)

To contextualize our findings, let's compare and contrast our themes to prior research. In his overview of the arsenic life controversy at NASA, Jogalekar (2015) concluded that the reviews he documented "may compellingly be said to be the very broadest form of peer review in terms of diversity, with scientists and readers from every conceivable field having access…" and "this informal peer-review process was liberating and democratizing…(Jogalekar, 2015)" In Science and Technology Studies (STS) parlance, this constitutes "accelerated virtual witnessing in the age of open science" whereby publicly processing scientific claims over the web leads to more rapid decisions about them (Meskus et al., 2018). Reviewers are indeed the motley crew of credential insiders and uncredentialed outsiders that we observed in our digital ethnography. In our study, we attended more to the informal reviewers by naming them, interacting with them, and revealing their cross-disciplinary nature.

In contrast to prior studies, we were able to describe authors' and publishers' clashes with informal reviewers more deeply and their proclivities to *hack homes by nesting and bridging* communities for extended interactions. This theme is our most unique one to contribute to this literature. It may simply be the case that niches developed by consistent informal reviewers were more difficult to find in other study windows. Given the current blossoming of social media protocols and applications, we expect that many more users will burrow into niches that organically evaluate research.

When describing a stem cell fraud case (STAP), Meskus et al., (2018) did not directly and succinctly state the *depth of reviewing activities*. They did, however, expose that informal reviewers scrutinized images in papers that may have been manipulated, found "inconsistent scale bars", images copied from a dissertation, and other matters beyond the failed replicability of the key results. To us, this suggests that our theme *intensive infrastructural inversion* has a precedent in the literature. Our digital ethnography is unlike the above studies, though, in that we catalog a wider range of reviewing activities and deepen our understanding of what it is like to do so work on a longer timeline.

Our fourth theme, *immunologic self-preservation* of authors and publishers, closely parallels the finding by Jogalekar (2015)[30], who remarked that for an erroneous chemistry paper on the purported synthesis of hexacyclinol, the retraction required six years. When even critical issues are uncovered, authors and publishers routinely deny, delay, and deflect. In our ethnography, unlike extant studies, we enumerated more ways in which authors and publishers coordinate tacitly to quell embarrassing news: retract and resubmit opportunities, excessively delayed retractions, and retractions with terse notices missing key contextual details.

The cases we presented in Results above are also unique in another respect; they span a spectrum of severity from minor to catastrophic. This was not coincidental. We intentionally selected them from a larger pool of cases to impress upon readers their diversity. Whereas case studies of informal reviewing have thus far focused on extended investigations of research integrity, we have attended to mundane as well as exceptional cases. Interestingly, our four themes typify the four norms or values of scientific practice conceptualized by sociologist Robert Merton: Communalism, Universalism, Disinterestedness, and Organized Skepticism (Merton, 1973). These Mertonian norms are a commonly used distillation of scientific values that are thought to mark science from non-science.

Communalism, or the *collective ownership* of scientific products and collaboration, appears in T1: motley crew) and T2: hacking homes by nesting and bridging. The informal reviewing community is open to scholars and laypeople from varied backgrounds, welcoming them without restrictive gatekeeping mechanisms. Users are free to comment in newly formed niches like Bluesky's Scientific Comments feed and the subreddit r/PeerReview, even if they lack credentials and exist outside of their discipline.

---

[30] Jogalekar, 2015, like us, noticed that across cases of informal peer review, public participation of diverse members rapidly uncovers problems in papers and scholarly revisions like retractions signal that their critiques are accurate.

Universalism, or the triumph of *impartial debate* over personal attacks and sociopolitical status isn't coded in our dataset and abstracted into a theme, yet we did note that explicit rules inside informal review communities like Bluesky's Scientific Comments feed and the subreddit r/PeerReview bar *ad hominems* and other forms of incivility. The moderator on Scientific Comments has stated as much telegraphically and directly. Unlike an editor at a journal or conference, the Scientific Comments moderator does not exercise power over a research report or its discourse directly. His only prerogative is the filtering of incivility and abuse. Considering that his technical infrastructure is so simple and capable of being copied by others, potential misuse could be counteracted by members who might create a rival feed.

Disinterestedness here refers to an orientation for *collective benefit* instead of personal gain. As we found in cases that revealed T4: immunological self-preservation, the infrastructure of formal peer review is often self-interested. Scholars, publishers, and institutions nominally strive to be disinterested, but falter because of incentives to shelter themselves in a self-spun knowledge matrix that advances shorter-term and self-serving ends of dominant scholars housed in common schools of thought.

Organized skepticism, or the ability to *critically evaluate* scholarly artifacts socially, is apparent within informal review infrastructures. Cases that gave rise to T2: infrastructural inversion, offered us a taste of the forensic scrutiny often missing from formal peer reviewing activities, such as probing open materials, preregistrations, and cross-checking publicly available datasets. In our observations and interactions with informal reviewers, we were regularly reminded of our experiences with and others' characterizations of formal peer review's disorganization. Informal review, at times, can be *more skeptical* than formal peer review and is powered by its own kind of organization, one that can be missed if presupposing unsanctioned evaluations as the anarchic complaints of firebrands.[31] Organization, for informal reviewers, is twofold. First, it summons underused and unglamorous cognitive-behavioral *processes* like checking links to open artifacts, statistical checks, and analyses of figures for fraud. Second, it wields software *technologies* in plain view to raise suspicions while cautiously hedging their judgments with conflicts of interest disclosures and requests to converse with other scholars. These skeptical checks on study reports are not absent from formal peer reviews, but we have seen that informal peer reviewers use them as if to tidy up after negligent formal peer reviewers.[32] Such tidying up is possible because of the greater number of reviewers per report and their superior judgements (Arvan et al., 2025).

These four themes, induced from the extended study of diverse cases, suggest that informal review embodies Merton's norms of science closely.[33] Whether informal reviewers can claim to embrace Mertonian values *more strongly* than formal peer reviewers is likely to stoke fierce debates. Future studies ought to test this possibility so that scholars don't argue from a favored and evidence-free point of view.

## 7.3 Strengths, Scope, and Limitations

We embarked on this project to study a topic, informal review, touched by only a smattering of empirical studies and a method, digital ethnography, rarely applied to metascientific questions.

---

[31] When focusing on isolated and extreme cases that polarize scholars and laypeople, it is possible, as with the Lancetgate case, to find a cacophony of voices eliciting the impression of "disorganized scepticism" (Schultz et al., 2024). In our ethnography, the efforts we took to sample diverse cases enabled us to witness organized skepticism; sleuths agreed multiple times within a single thread, assisted each other, and coordinated letters, appeals, tools, and next steps.

[32] In future work, it may be fruitful to categorize the type of work that formal and informal peer reviewers do into levels. We know, for instance, that formal peer reviewers and editors do not check pre-registrations against manuscripts for consistency between study plans and implementation of the plan (Syed, 2025).

[33] Scholars in this area studying the impacts of sleuths have made the same connection (Yeo-Teh & Tang, 2023).

To answer our questions, we did not simply spotlight individual reviewers, settings, practices, or technologies. Instead, we adopted a systems-level or infrastructural lens through which we described the assemblage of people, settings, practices, technologies, and norms. The empirical studies we reviewed in our Literature Review (§3) above lacked this infrastructural lens. When they framed their inquiry narrowly, they missed the opportunity to capture their synergistic interactions.

Our ethnographic epistemology aimed for an insider's perspective of a culture in its natural context and laid bare the rationales, full critiques, and stimulating discussions that a positivist, look-from-afar study would have missed. Soon into the study, we found ourselves data-laden and needed to triage cases for analysis. We stumbled upon behaviors unexpectedly and were empowered as insiders as we mirrored reviewers' behaviors. The generative power of digital ethnographies is well-suited for discovering insights, especially when conducted longitudinally. Timing data sourcing during 2024 serendipitously benefited us in being able to observe the Academic Twitter migration. Infrastructural breakdown, invention, and maturation unfolded without our intervention and will likely be worth researching for many years longer as new members join, technologies evolve, norms shift, and the avant-garde becomes the old guard.

When scoping our study, we restricted ourselves to English-language posts on popular platforms over a twelve-month timespan and eschewed formal and quasi-formal reviewing initiatives. For instance, we did not analyze discourse from semi-formal review organizations like ASAP Bio PreReview club, repliCATS (replicating Collaborative Assessment for Trustworthy Science), and Peer Community In (PCI) despite membership in these communities. When we began our data sourcing, we were only aware of social media sites and PubPeer as the most active venue for public discussions of research reports. A few other communities like Peer Review (http://peerreview.io/) and Octopus (www.octopus.ac) were used infrequently and didn't merit our attention. Our insights and themes, moreover, do not speak to the emerging importance of automated research evaluation technologies using large language models (LLMs) in public and private digital communities. A more complete description of informal review and its dependencies requires more attention to what such technologies permit and deny.
Our attempts to sample theoretically from the larger pool of informal review cases relied on memory and judgments of skimmed notes. It was not as thorough as we would have liked. We do, however, consider each of the four themes from Insights above to be enlightening themes that are worth sharing with the broader scholarly community, capable of being reinforced by follow-up studies, and amenable to brainstorming practical solutions.

## 7.4 Contributions

As we expanded our understanding of informal review, we extended the literature with empirical, ontic, theoretic, and practical tracks. After our long immersion into this literature and community, we realized the need for a single report synthesizing the nature and tooling underlying informal peer review. Doing so, we felt, can hasten reforms and infrastructuring. In section 7.2 above, the description of Triangulations and Interpretations is the same as our empirical contributions. In the next section, we describe the ontic contributions. These are essential for understanding our theoretic contributions, which in turn will guide our practical contributions.

### *7.4.1 Ontic Contributions: You are the peer*

“What is common knowledge in your field, but shocks outsiders?
We're not clear on what peer review is, at all” (Overly.Honest.Editor, 2025).

“The term "peer-review" should be expanded to encompass various experts and platforms beyond traditional academic journals” (Irawan et al., 2024).

What, then, is a fitting and distinct name for this phenomenon? As we described at the outset of this paper, we struggled to name it. During our digital ethnography, we toyed with names like informal critical appraisal, swarm critical appraisal, informal review, unbundled peer review, scholar-led peer review, and open peer review. After much reflection and debate, we realized that the variant of “open peer review” that we are interested in is a subtype of open peer review that blends open participation (aka “crowdsourced peer review” wherein diverse agents can review), open interaction (meaning author-reviewer and reviewer-reviewer discussions can occur unmediated by others), and open platforms (aka “decoupled review” wherein organizations and technologies to facilitate reviewing exist and operate independently) (Ross-Hellauer, 2017). In *our* variant of open peer review, reviewing can occur anonymously and doesn’t require preprinting. Informal review can occur before publication as pre-publication peer review or after publication as post-publication peer review. We observed both kinds in our study and appreciate each one.

After conducting this digital ethnography, we concluded it would be productive to use the term *informal peer review,* just as other scholars have done so (Butchard et al., 2018; Gowers, 2017; Müller et al., 2025; Yeo et al., 2017). To the feature *open participation*, we add organicity of how the work initiates. To the property *open platforms,* we add that platforms are most often free, intuitive, and interconnected. This form of reviewing is typically organic, composed of diverse agents, inclusive, free-to-use, and distributed across platforms. See Table 3 for a synthesis of informal review's properties.

**Table 3.** Key properties of informal and formal peer review: participation, interaction, and platforms.

| Properties | Informal Peer Review | Formal Peer Review |
|---|---|---|
| *Participation:* Who is the peer? | work emerges bottom-up from reviewers (always) | centralized, small, ad hoc team, private; selected experts review (always) |
| | crowdsourced and public; diverse agents review (often) | |
| *Interaction:* How (in)formal is reviewing? | interactions between authors and reviewers as well as between each reviewer are unmediated by publishers (always) | interactions between authors and reviewers as well as between each reviewer are mediated by publishers (always) |
| *Platforms*: Where is reviewing? | reviewing uses publisher-independent organizations and technologies (always) | reviewing uses publisher-dependent hierarchy and technologies (always) |
| | platforms are free to use, interconnected to other digital communities (often) | platforms are paid and isolated from other digital communities (often) |

Formal and informal peer review adopt incarnations and subsume a blend of these traits. Considering our cases above, one might want to conceptualize each property on a continuum rather than as a binary. Levels of open participation, interaction, and platforms may exist. So to conceptualize informal peer review more generally, we analyzed seven cases from Results above (§6) and reported a summary in Table 4.

**Table 4.** Selected cases of informal peer review and their properties help abstract its unique nature.

| | **PARTICIPATION** | **INTERACTION** | **PLATFORMS** |
|---|---|---|---|
| **Cases** | **Diversity** | **Informality** | **Independence** |
| 1. Economics degrowth review | cross-disciplinary scholars | unmediated by authorities | Twitter & PubPeer: independent |
| 2. Re-analysis of social media | cross-disciplinary scholars, includes Ph.D. student | unmediated by authorities | Twitter & PubPeer: independent |
| 3. Incorrect power analysis | cross-disciplinary scholars, includes retired professor | unmediated by authorities | Twitter & PubPeer: independent |
| 4. Names shape faces | cross-disciplinary scholars | unmediated by authorities | Twitter & PubPeer: independent |
| 5. Metascience preregistration | cross-disciplinary scholars | unmediated by authorities | Bluesky & PubPeer: independent |
| 6. Cassava Sciences fraud | cross-disciplinary scholars, investor | unmediated by authorities | Twitter, PubPeer, blogs, sites: independent |
| 7. Apple cider vinegar | cross-disciplinary scholars, sleuth | unmediated by authorities | Twitter, PubPeer, blogs, sites vs. BMJ Rapid Responses |

From these cases, we gained confidence that the term *informal peer review* captures the open participation and interaction on open platforms in our cases. Their unique properties are worth minding: organicity in the origin of work, distributivity across publisher-independent platforms, inclusivity of participation, and the diversity of reviewers.[34] From them, we propose a working conceptual definition. Informal peer review can be defined in many ways and we reject the essentialist assumption of one right definition for constructs (Wittgenstein, 2009). Instead, we find it valuable to consider definitions as *tools* that focus attention on phenomena and entities (Howat, 2024). For our purposes, it is *pragmatic or productive* to define informal peer review as:

> the practice of evaluating research inclusively on publisher-independent platforms with direct interactions between authors, reviewers, and outsiders; often self-organized and may cause revisions like corrections, notices of editorial concern, and retractions

By "informal", here, we mean that such reviews are typically unofficial and unrequested. Informal peer reviewers lack the weight of quality control that formal peer reviewers exact on submitted manuscripts. To some, an "informal" review may carry a negative connotation and connote the possibility of unprofessional behavior. That is not our intent and grossly mischaracterizes the bulk of informal peer reviews that we laboriously documented over the course of this ethnography. We chose to use "informal" for its ability to characterize the casual behaviors that we observed.

---

[34] In a busy social media thread from an open science reformer seeking a more neutral term for researchers studying research integrity issues, we noticed many commenters disagreeing about the ideal term. To our ears, even options like "sleuths", "science auditor", "validator" "forensic scientists", and "post-publication reviewers" seemed clunky and unpalatable for the broader scholarly community. We do occasionally refer to persistent informal reviewers as *sleuths* to distinguish them from occasional reviewers. Even long-time practitioners and open science scholars disagree on what to call this practice. As one long-time sleuth noticed, "There is no encased term for a scientific critic." and paraphrasing a Game of Thrones line, "A critic has no name" (Heathers, 2018)." In the early years of the research reform movement or Replication Crisis, traditionalists branded the most vociferous critics targeting serious issues in reports to be "online vigilantes", "destructo-critics", "data thugs", "methodological terrorists", "self-appointed data police", and other unflattering epithets (The Observer, 2016). Online commenters have proffered alternative titles like "science critic", "integrity enthusiast", and "critical meta-scientist," though we find that none of these optimize precision, tone, and brevity to our liking or that of others. For now, a sleuth has no name.

Our use of "peer" has provoked controversy among some scholars, given that reviewers are cross-disciplinary and at times exist outside of academia as investors, hobbyists, and anonymous agents who may confess to lacking a scientific background. Nonetheless, their achievements are unique and impactful, rivaling traditional in-group members who are typically called peers. In cases like Cassava Sciences fraud (§6.2.3 above), we've observed that the coordinated efforts of traditional scholars and nontraditional commenters surfaced serious academic and industry fraud, resulting in a cascade of widespread business and scientific aftershocks. Given their accuracy and volunteered contributions to the scholarly corpus, we welcome them as peers.[35] Our pragmatic outcomes-based approach leads us to "settle the 'peer' question... is peer commentary the same as peer review? (Harnad, 1998)" in the affirmative.

The word "review" here may also confuse many, so we wish to clarify that we are using it more generally than a formal peer review report. Here, we mean "a critical evaluation (as of a book or play)" rather than the sense of a comprehensive and regulated peer review for a conference or journal (Merriam-Webster, 2025). Typically, formal peer reviews are expected to be structured with a summary of the manuscript and a lengthy list of critical remarks followed by a list of specific comments referencing specific content in the manuscript. These formal peer reviews, when implemented according to community norms, can be lengthy and written in an academic style. This is not the case with the often brief, casual, and passionate informal peer reviews. Just as we reject the narrow scope of peer in peer review, we claim that the narrow use of *review* denigrates the impactful work that informal reviews do across online communities. Rather than attend to superficial features of reviews like their length and style, we favor attending to their intentions and impacts. To the old question "... is peer commentary the same as peer review? (Harnad, 1998)" we answer that it is productive to consider it to be so. Let's not denigrate insightful knowledge work to mere commentary, as it carries a connotation of uselessness and passivity.

---

[35] Other teams of metascientists and STS researchers, working independently, have likewise concluded that their work "implicitly challeng[es] what constitute a 'peer' for the purposes of peer review, extending the definition in various ways (e.g., specialist statistical reviewers, impact specialists, expert patients, or researchers working in non-academic settings), making it important to take these into account (Waltman et al., 2023)."

Informal peer reviews can be posted in public venues, like the websites and social media platforms we described or in private venues like Slack channels, Discord servers, Mastodon servers, and invite-only subreddits. Because of privacy policies, informal peer reviews in private settings are likely to remain the mysterious dark matter of this universe. One might refine this definition in future studies by studying this phenomenon at scale and rating cases on all three properties.

From our interactions with colleagues and study members in our ethnography, we learned that many perceive such reviewing "with only the negative 'yin' of criticism, correction, retraction, and failed replication (Bastian, 2014). To do so would miss the valuable complement, or "the 'yang'...Answers to questions may be critical for other studies, for adequate research assessment and synthesis, and for considering practice and policy implications. Discussion can build, apply, connect, and update ideas... (Bastian, 2014)" After understanding the meaning of informal peer review, we advise studying it *oneself* before tinging it with undeserved connotations.

Constructs like informal peer review and names to reference them are shaped using scholarly intuition and could benefit from comprehensive evaluations according to shared criteria. Eight such criteria of good concepts are available to us: familiarity, resonance, parsimony, coherence, differentiation from related constructs, depth, theoretical utility, and field utility (Gerring, 1999). We argue that the moniker informal peer review, as with other constructs in the sciences, satisfies some, but not all, of these criteria. One finds tradeoffs in their construction. Nonetheless, informal peer review is *familiar* in that others have used the term in literature. It is *resonant* inasmuch as it is memorable, *parsimonious* because it limits itself to subsuming just three features of open peer review and still covering an expansive array of cases, *coherent* to our knowledge, clearly *differentiated* from formal peer review as its antithesis, and *deep* enough to study at length. As we have seen now, our ontological labors have helped organize the conceptual tangles with post-publication peer review and open peer review. We now enjoy a one-to-one relationship between the construct informal peer review and its name – *field utility*. As we will show next, the phenomena we gather under the construct informal peer review also unearth systemic properties and problems worth investigating – *theoretical utility*.

### 7.4.2 Theoretic Contributions: The patchwork works

“Post-publication evaluation is highly fragmented. It often appears within future articles… or in formal research syntheses (Bastian, 2014).”

“In a patchwork, scholars access work here…read commentary there… [The] System is the attempt to overcome lack of shared practices (Elfenbein, 2025).”

"There's no organization, there is no structure, there is no backup, no retirement plan, no money, no incentives, and… not even… agreement… (Bartlett, 2018)"

Before data sourcing I (JP) noticed a shortage of explicit, evidence-based theorizing about informal peer review using diverse cases in which reviewing provoked disparate impacts. Rich observational research, social scientists remind us, can form the lattice upon which sturdy descriptions, explanations, and predictions can accrete (Corker, 2021). Here, we attempt to do just that by theorizing informal peer review and formal peer review together from a systems-level or infrastructural point of view. From this view, we ask a broad, descriptive question: What is the infrastructure of peer review systems like?

The germ of our new descriptive theory begins with the observation that reviewers' scrappy improvisations with the resources available at hand help them develop a fragmented bricolage or *patchwork evidence infrastructure* more like a beaver’s dam or a bird’s nest made from ready-at-hand resources instead of bespoke materials architected with first principles design.[36] We find it productive to define patchwork evidence infrastructure[37] as a subtype of an evidence infrastructure (Harvey, 2016):

---

[36] This style of development, hacker culture, is reminiscent of the jugaad type of frugal repair and innovation on the Indian subcontinent. Just as jugaad inventions like makeshift vehicles and electronics use existing resources because alternatives are unavailable, informal reviews resort to using existing, less-than-ideal resources. Informal reviewers nest in general-purpose platforms, instead of developing technologies and communities de novo.

[37] For earlier reference to the ideas underlying patchwork infrastructures, see Crabu & Magudden, pg. 154 on inverse infrastructures and other bottom-up infrastructures.

> the fragmented assemblage of ready-at-hand, reviewers, communities of interest, and technologies that evaluates and shares research reports in public settings; often diverse, resource-poor, and self-organized

In this section, let's apply our construct patchwork evidence infrastructure to theorize about informal and formal peer review[38].

To understand the sociotechnical *structure* of informal peer review, we asked: Who reviews and where? Recall from T1: motley crew that informal peer reviews are a motley crew or *a patchwork of people*[39] who assemble in one or more digital settings often by happenstance. Also recall that the setting of informal peer review is a network of niches instead of an all-in-one homebase with all the desired features and funding required to complete the work (T2: hacking homes). In contrast, formal peer review operates under the banner of credentialed experts in specific topic areas using bespoke reviewing software developed and maintained by publishers and vendors with stable income.[40]

To understand the *functions and mechanisms* of informal peer review, we wondered: Why and how do reviewers review? Reexamining the governance theory of peer review provides an answer (Reinhart & Schendzielorz, 2024). It theorized three key functions for formal peer review – quality control, distribution of scarce resources, and self-governing science away from external interference. We think that these functions describe journal and conference peer review as well as grant peer review. Informal peer review, though, does *not* distribute any scarce resources. Instead, it seems to function to decide who gets to *keep* scarce resources like space in publications and grant funding.

When we consider how informal peer review engages in quality control, we recall T3: intensive infrastructural inversion. It states that informal peer reviewers use cognitive and technological tools to deeply evaluate the credibility of claims and evidence. Although skeptical reviewing such as this is not necessarily restricted to informal peer review, we noticed that informal reviewers often cleaned up after the mistakes of formal peer reviewers. Reviewing as part of journals, conferences, and grant panels seems to lack the depth required to ferret out fraud, plagiarism, undisclosed conflicts of interest, conceptual fumbles, and analytic errors. From this, we gather that the functions and mechanisms of informal peer review overlap greatly with those of formal peer review while diverging in their diversity and depth. Occasionally, journals and grant funders may impel formal peer reviewers to use checklists and quality appraisal tools to guide their evaluations, though no such enforcement exists for informal peer review (Speich et al., 2020).

---

[38] Here, formal peer review subsumes journal peer review, conference peer review, and grant peer review.

[39] Perhaps the term "flash team" from the team science and human-computer interaction literatures also applies here, as small teams of reviewers can form and evaporate quickly (Retelny et al., 2014).

[40] This stability, one often finds, does not guarantee a satisfactory user experience.

To understand the *behaviors* of informal peer review, we asked: What happens after reviewing among the triad of authors, editors, and publishers? As T4: Immunological self-preservation exposed, the triad supports each other amid a cascade of reputational impacts by coordinating a patchwork of defensive maneuvers: delaying, denying, and downplaying scholarly issues. There seems to be a playbook for handling accusatory informal peer reviews in the Committee on Publication Ethics (COPE) Guidelines, but publishers apply them only loosely to the irritation of informal peer reviewers (Committee on Publication Ethics, 2025). After journal and conference reviews are distributed for responses, authors largely oppose critiques from editors and publishers. Crossing the threshold of publication bonds these previously rival forces into a likeminded triad invested in safeguarding the publication. The dynamics differ in the case of grant peer review, as there is less contact between proposal authors, editors (program officers), and grant panelists.

We did not initially ask ourselves: How does informal peer review *legitimate* itself? This question was posed to us by colleagues who reviewed an early version of this report. On reflection, we felt this question consequential enough to merit a response. It also reminded us of peer review theorists' questions and analyses about legitimacy (Hug, 2022; Reinhart & Schendzielorz, 2024).[41] It appears, from the Cassava Sciences fraud case above, that informal peer reviewers do not have the legitimacy of formal peer reviewers whose powers filter problematic manuscripts from publication. Instead, informal peer reviewers require legal support to dissolve the powerful bonds between authors, editors, and publishers. In the eyes of many scholars and study authors, informal peer reviewers are unwelcome, obtrusive, and illegitimate. They do not wield the credibility-attracting regalia of formal peer reviewers: blinded review, ostensibly neutral editors, established interaction policies, rubrics, checklists, and the long-seeming history of credentialed experts sanctioned to ritualistically anoint manuscripts as trustworthy.

[41] Peer review's legitimacy, which is useful for decision-making and accountability, relies on experts, deliberation procedures, and a written record (Reinhart & Schendzielorz, 2024).

Recall that in the case Metascience Preregistration above, at least one of the informal reviewers was also a formal peer reviewer. Issues discussed during informal peer review like the missing preregistration of key claims were preceded in formal peer review. Curiously, the study was later retracted with a resubmission opportunity for issues apparent during formal peer review. This may constitute a rare case of informal peer review being functionally *more* legitimate than formal peer review in the eyes of the *Nature Human Behavior* editors. Both informal and formal peer review leverage a patchwork of methods to legitimate themselves, though we are presently in the dark about the full suite of techniques favored by informal reviewers. In our ethnography, hints of a toolkit are evident: viral posts across social media, communicating with authors in public settings, and making evidence-based accusations in courts. Which other legitimizing techniques exist?

As a whole, informal peer review operates in so threadbare an evidence infrastructure that it belies its efficacy. If it were a government, it would be a *stateless anarchy* in the positive sense of the term. No formal government exists to rule its members and adjudicate disputes, but the people operate in ways that occasionally establish basic norms to minimize potential harms (see Bluesky feed Scientific Comments and the subreddit r/PeerReview). Informal peer reviewers mentor others, make recommendations, share tools, and admonish miscreants. Unlike many alternative forms of governance, anarchies carry an undeserved negative connotation. They may, contrary to widespread thought, operate positively with largely harmonious relations. Given that informal peer review occurs publicly online, many voices can speak. This open participation is democratic with weak central enforcement over the discourse. Nonetheless, a communal rule of law emerges. See Table 5 for an overview of governance models of informal and formal peer review.

We postulate that formal peer review, operating under publishers and funding agencies, is more akin to a *pseudo-meritocratic oligarchy*. This governance structure recruits a band of credentialed experts in a topic area to enact quality control and distribute scarce resources like publication space and funding. As we have witnessed, this structure is not without its limitations in even the most elite settings. Because reviewers and panelists are not elected by the scholarly community as a whole, formal peer review is not a direct democracy. In some settings, study authors can nominate reviewers, but it is unclear the extent to which their nominations succeed in electing their desired reviewers.

Informal and formal peer review, we have concluded, reside on polar extremes of *infrastructural maturity*. Whereas one is viewed as legitimate, the other works overtime to be recognized. One enjoys status, tooling, norms, and rapid impacts in quality control, whilst the other must construct these from scratch. They resemble one another only in their commitment to upholding the Mertonian CUDO(S) norms.

Sketching the maturity of an evidence infrastructure this systematically can be revealing, especially when integrating elements like incentives. Equipped with this lens, we might more easily admit an uncomfortable truth –we are demanding that formal and informal peer review do more than they *can* do. Though formal peer review enjoys greater infrastructural maturity, it is by no means a comfortable home for many scholars. That work is also volunteered, performed in the crevices of one's calendar, lacks widespread training opportunities for new entrants, and could benefit from greater procedural guardrails (Willis et al., 2023). It is no surprise that reviewers feel burdened and that reforms are underway. Whichever variant one cares for, key resources are required from their inception. When predicting the success of scholarly work more generally, we wonder: Which resources are available to busy scholars to support this work?

If an asymmetry exists between the task demands and the infrastructure supports, each member of the community will perpetually play catch-up. It is worth minding that both formal and informal peer review emerged organically, as incidental responses to broader sociotechnical forces. Their origins were unplanned, trajectories unknown, and efficacy initially unexamined. As such, their designs and operations could be fundamentally ill-suited to accomplish their many functions.

**Table 5.** Analysis of informal and formal peer review according to their governance structures and infrastructural elements.

| | **Informal Peer Review** | **Formal Peer Review** | |
|---|---|---|---|
| | | Journal Peer Review | Grant Peer Review |
| *Structures:*<br>• who and where | T1: motley crew<br>T2: hacking homes | credential experts<br>bespoke software | credentialed experts<br>hybrid settings |
| *Functions & Mechanisms* | | | |
| • why (process: quality control) | quality control and error correction | gatekeeps or filters reports perceived as lower-quality and/or less relevant to venues | |
| • how (process: depth of reviewing) | T3: intensive infrastructural inversion uses a patchwork of common and uncommon tools to evaluate reports | | |
| • why (outcome: distribute scarce resources) | doesn't distribute scarce resources, but does control quality | distributes scarce publication space | distribute scarce grant funding |
| • how (process: constructive feedback) | gives constructive preprint feedback, but post-publication is too late | gives constructive feedback[42] | gives constructive feedback |
| • why (context: self-govern science) | reviewers may be scholars or non-scholars; both supports each other's efforts to govern science | created to manage editorial burden | evolved to self-govern science from political intrusion |
| *Behaviors*<br>• what happens? | T4: immunological self-preservation of author-editor-publisher triad | authors, editors, managers, and others compete to implement their own goals | |
| *Legitimacy* | Triangulation of critiques, legal support, scholarly revisions | credential experts, blinded review, editors, policies, long-seeming tradition | |

[42] In discussions of peer review, the "…function is providing feedback… whether or not it is accepted… you receive valuable advice. (Gowers, 2017)" Histories of formal peer review don't describe any *a priori* intentions to use reviewing as feedback, so we consider it to be an evolved function.

Resting upon this more solid descriptive theory, we asked ourselves a normative question: How *should* peer review systems be governed for responsible and effective evaluation at scale? Inspiration arrived from two other peer production communities: Wikipedia[43] and the open-source software (OSS) community that develops the Linux operating system kernel. Wikipedia and OSS communities mirror informal peer review in that they support open *participation* by interested volunteers, open *interaction* minimally mediated by authority figures[44], and open *platforms* Wikipedia.org and GitHub on which creative knowledge work moves. Like the informal peer review community, these peer production communities invite a diverse crowd of contributors whose overlapping expertise, interests, and schedules unlock productivity and creativity.

In their early years, these peer production communities shared a high degree of bricolage or patchiness in the form of variegated and ready-made groups, materials, settings, and social pacts that arose bottom-up to resolve immediate, pressing problems. Wikipedia was considered a "creative anarchy" hamstrung by laissez-faire leaders and trolls with a passion for flaming (Poe, 2006). To realize what we mean here, consider rival systems like formal peer review*, Encyclopedia Britannica*, and proprietary software development teams with their planned, top-down leadership using vetted experts implementing routine procedures. The resources available to them, project development, and outputs are of an entirely different kind. The most structured forms of peer review, FDA Drug and Vaccine Review, and IPCC committees, are the least patchy as they benefit from significant funding and status in the eyes of others (see Fig. 16).

---

[43] Scholars yearning for alternatives to formal peer review and journal-led publishing offer much (Cummings, 2020).

[44] Wikipedia is ultimately dependent on its laissez-faire co-founder Jimmy Wales and the Linux community has a "benevolent dictator for life (BDFL) in founder Linus Torvalds.

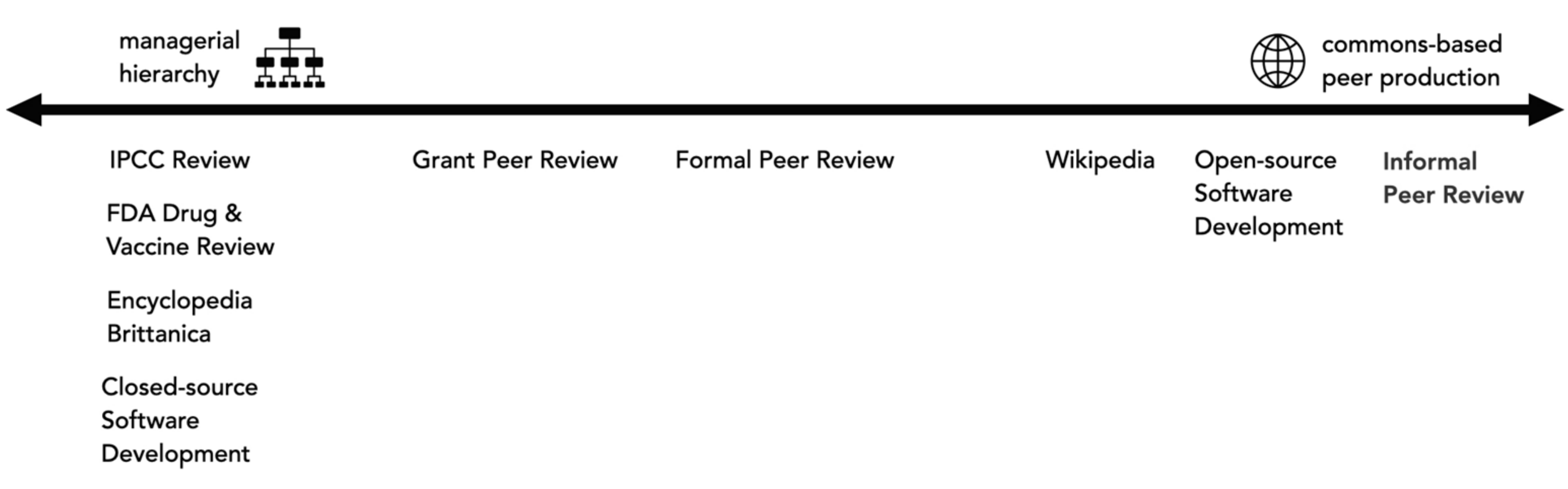

**Figure 16.** Our proposed spectrum of governance across peer production communities in scholarship and daily life spans higher levels of governance as in a managerial hierarchy (left) to lower governance as in a fully commons-based peer production community (right). We concluded that informal peer review more closely resembles commons-based peer production systems like Wikipedia article authoring and open-source software development.

These evidence infrastructures approximate the managerial hierarchies of corporations instead of the flatter, meritocratic peer production communities. Whereas managerial hierarchies are selective about who counts as an insider, secretive, authoritative, and use phased (waterfall) project management, peer production communities are permeable about their membership, solicit contributions from legions of unpaid volunteers, and believe that Shipping beats perfection. This agile approach to knowledge work, which updates products quickly, is especially useful when facing complex problems.[45]

Soon after their rise, some peer production communities bore damages that stimulated reforms. Wikipedia's skirmishes with trolls and self-serving users stimulated its organic transition from a flat, creative anarchy to an increasingly hierarchical and bureaucratic machine where discussions of governance occurred more frequently. Co-founder Jimmy Wales watched over Wikipedians, but intervened only to clear rare and serious impediments to cooperation. These governance reforms roughly paralleled technological reforms like semi-automating spam and vandalism detection. The OSS community developing Linux, largely due to the personal technological needs of its founder, also transformed organically from a team-of-one to a large-scale, meritocratic-authoritarian model in which a laissez-faire leader oversees numerous maintainers regulating parts of the Linux kernel (Moody, 1997; Torvalds & Diamond, 2002). The best perceived code contributions or "patches" are used and contributors must resolve disagreements about which of two conflicting patches to use.

Particularly impassioned informal peer reviewers have been critiqued for their occasional incendiaries and fumbles (Bartlett, 2018). When they use unvetted forensic methods to undermine the reliability of evidence, readers begin to reconsider their judgements about a study. As a peer production community without an ultimate leader, there are few visible signs that informal peer review has *significantly* evolved its core governance, despite calls for them (Darda et al., 2023; Derksen & Field, 2022; Whitaker & Guest, 2020). One emerging exception is the training manual and norms guide COSIG (Richardson, 2026). See Fig. 17 for an overview of governance trajectories.

---

[45] In software engineering and the open-source community, one finds a contrast between top-down design of "cathedrals" and bottom-up design of "bazaars" (Eghbal, 2020). We begin by using "managerial hierarchies" instead of "cathedrals" and "peer production communities" in lieu of "bazaars" to connect to our key social science constructs commonly used in academic literature. Tellingly, Eric Raymond's influential essay and eventual book *The Cathedral and the Bazaar: Musings on Linux and Open Source by an Accidental Revolutionary* partially influenced Jimmy Wales to develop Wikipedia (Poe, 2006).

| (pure) commons-based peer production | Birth | Maturation |
| --- | --- | --- |
| informal peer review (stateless anarchy) | - individual-level norms; formal policies<br>- solicits guidance from their network | - norms and training guide (COSIG) in development now* |
| open-source software development (meritocratic laissez-faire monarchy) | - BDFL regulates with consensus<br>- self-selected contributions<br>- minimal governance documentation | - tiered commit access: Apache, Linux<br>- formal documentation and processes<br>- security teams; structured releases |
| Wikipedia (constitutional consensus parliamentary democracy) | - "Be bold"; minimal criteria for notability<br>- flat editing hierarchy; open contribution<br>- ad hoc talk-page dispute resolution | - Five Pillars; policies for notability<br>- ArbCom; admin elections (RfA)<br>- sockpuppet detection, article tiers |
| formal peer review (meritocratic oligarchy) | - editor-selected referees; informal stds.<br>- single/double-blind reviews<br>- editorial oversight for accountability | - COPE guidelines; checklists emerging<br>- Registered Reports<br>- open (public) peer review |
| (pure) managerial hierarchy | | |

**Figure 17.** Managerial hierarchies like formal peer review and peer production communities like Wikipedia, Linux, and informal peer review have evolved diverse forms of governance including guidelines, training manuals, formal procedures, tiered access, values, policies, etc. Informal peer review currently lacks the comprehensive governance apparatus of other production systems.

Within formal peer review, governance transpires from membership in the community or *insider status*; we trust that peer reviewers are in fact peers because they are invited by community-elected editors or appointed based on reputation. Standards of review, too, can be enforced using society codes of conduct or reviewing standards and criteria at journals and conferences, enforced with a range of formal mechanisms and sanctions. To the questions of governance and legitimacy raised by our investigation here, we suggest that informal peer review could draw on the rich design space of governance strategies used by Wikipedia to preserve the benefits of open participation, interaction, and platforms while mitigating its harms. In some cases, restrictive algorithmic quality control mechanisms have succeeded (Geiger, 2017; Halfaker et al., 2014). Though tempting, requiring users to create and use accounts to post may constrain the growth of communities without enhancing quality (Benkler, 2006). Onboarding newcomers in structured, personal ways may prevent attrition and carefully designed mechanisms to revert changes can increase the quality of contributions while balancing openness (Halfaker et al., 2011; Morgan & Halfaker, 2018).

If the analogical bridges between informal peer review, Wikipedia, and OSS development are justified, then informal peer review may also need to explore how much to formalize its operations just as Wikipedia now carefully balances a flexible and evolving "bureaucracy" and governance structure as it invests in strategies to promote participation from newcomers. It is unclear whether and how these strategies would work for the motley crew of informal peer reviewers. One may be tempted to simply transfer in what works from one setting to another, but we caution that doing so may be futile because the scholarly community and its missions are not entirely like volunteer writers or software developers. Precisely how to proceed without losing informal peer review’s unique strengths is a compelling puzzle worth resolving with implications for the design and redesign of informal peer review platforms.

### *7.4.3 Practical Contributions: Discover, aggregate, and solicit*

“Substantive discussion in journal clubs, in email lists, in social media, and… conferences are not distilled into a concise, permanent, accessible record… (Bastian, 2014)”

"Science needs better defense mechanisms: researchers, journalists, and the public deserve tools to distinguish scientific rigor from its imitations. (Silvi & McIntosh, 2025)"

Our digital ethnography and accompanying conceptual contributions above suggest some immediate practical implications. They are productive for resolving a recalcitrant and "wicked problem" — how to address the information needs of readers and science “publics”. One may, proceeding cautiously, be able to mature informal peer review’s sociotechnical infrastructure to the point that it scales up to meet the needs of all. In this section, we begin with a systems-level diagnosis of informal peer review infrastructure, prescribe remedies, and anticipate potential downstream contingencies.

At their simplest level, peer review evidence infrastructures can be characterized according to its *size*. We think of formal peer review as being akin to a construction crew led by a supervisor and capable of constructing grant "cathedrals" according to fixed plans and schedules; like a construction project, delays are common. As a consequence of its historical primacy, there are many more formal peer reviews than informal ones. Millions of scholarly reports are being reviewed annually now and that number is growing despite reports of "reviewer burden." This is not so with the massive hubs, or what we call "megachurches", where informal peer review occurs. At PubPeer, one unpublished estimate reports that users have collectively commented on just 0.33% (250,000) of indexed Scopus publications across all years and disciplines (76,500,000) (Gierschner, 2025). Even long-time and trusted hubs like PubPeer contribute too few reviews to overtake formal peer review.

Informal reviews appear to be more common *outside* of megachurches like PubPeer; they are most visible in the buzzing bazaars of social media sites and personal websites. To estimate their prevalence crudely, we searched Altmetrics Explorer for Twitter posts over 2025 (Digital Science, 2026). Out of 182.5 billion total posts, there were 11,112,160 research mentions or 8,899,262 individual posts. This means that just 0.00006% of Twitter posts are research mentions and 5.8% of research mentions were negative. Estimating the exact proportion of these critiques that are *substantive critiques* is worth investigating as well as more accurate methods of classifying posts by sentiment, given the low accuracy of Altmetrics’ sentiment classifications. Using Altmetrics Explorer again, we found that 0.35% of all Bluesky posts in 2025 commented on specific published research reports, about 58 times more than Twitter (Digital Science, 2026). In raw numbers, this means that from 1.41 billion total posts, 4,979,775 mentions or 4,887,719 individual posts mentioned research. This is quite respectable for a nascent platform just beginning to unfurl its wings. Of these, Altmetrics labeled 170,309 mentions or 168,772 individual posts as negative sentiment because they were thought to critique research. These negative sentiment posts constituted just 3.3% of the overall dataset. In future, we expect that scholars will estimate the distribution of research mention posts across social media sites more accurately.

We consider it fruitful, given the cases described above, to *raise the frequency* of informal peer review. Doing so can help correct the scholarly record, surface valuable connections between literatures, reorient scholars to more productive ends, and enhance our understanding of research methods. We were not surprised that informal peer review is less common. Formal peer review emerged due to systemic threats to science, received systemic support from insiders, and has an eventful history. In this early digital era, we are just beginning to engineer the incentives and other mechanisms that will germinate informal peer reviews. Problems that plague informal peer review, like scale, somewhat affect formal peer review too. Consider, for instance, that systematic reviewers who find problems in randomized controlled trials (RCTs) do not have a simple mechanism to alert authors about their findings. Formal post-publication peer review in journals has "demonstrated limited effectiveness in identifying methodological and reporting issues. (Davidson et al., 2025)." No matter the variant of peer review that we attend to, there are opportunities to increase the solicitation of critical feedback.

Informal peer reviewing, we have found, is not the most appealing practice to discuss with broad audiences. The workers and their impacts are consequential but perceived negatively. Given this, we think that informal reviews should be supported with greater care towards branding. In many cases, terms like review, evaluate, critique, and similar ones should not be used. Even the most successful communities receive far fewer informal peer reviews than needed. Success will arrive when community builders and tool developers sugarcoat the practice with the sweet or neutral taste of other, more palatable activities. Platforms ought to focus more on soliciting and relaying feedback. In parallel, one might Trojan Horse the practice of informal review into the broader practice of synthesizing research. Informal reviewers and their champions would do well to recruit a community for a mission broader than merely reviewing papers, a mission that many find uninspiring.

From a more distant vantage point, the key needs for a successful evidence infrastructure become visible. There are *hard resources* like hardware, software, and data as well as *soft resources* like incentives, time, human contributors, socio-organizational structure, legitimizing mechanisms, and training. Informal peer reviewers, like formal peer reviewers, could benefit from many of the same resources, as both encounter patchiness to varying degrees.

Protected time, financial compensation, and quality training are hard for both kinds of reviewers to find. For informal reviewers, as the work is lower status, the constraints are felt more deeply. Error correction and fraud detection are not expected or welcome contributions on CVs and resumes. Scholars are not only disincentivized from flocking to informal peer review mega-churches, but they are also disincentivized from publicly critiquing in-group members. According to one post-mortem analysis of the failed platform peer-review.io, "People respond to people, not to abstract 'opportunities' and volunteer/service energy isn't infinite. A lot of the systems we have - including journals - are designed to work from that fact." (Heard, 2024/25).

On PubPeer and social media alike, *user friction* is also a pernicious problem. If a user reads a report and wishes to leave a comment, they must traverse a trail on which each subtask thwarts progress. Users must:

1. Decide whether their comment is accurate
2. Decide whether their comment is potentially valuable
3. Decide where to leave their comment
4. Visit the venue
5. Learn the norms and technical requirements of the venue
6. Connect their comment to the specific report and author it
7. Submit the comment, itself a tricky matter if using journals or PubPeer

On this trail, it is easy to lose interested and insightful contributors. As one physiologist remarked on Bluesky:

> …I want to get better at… giving feedback on preprints (both constructive and plainly positive). There are lots of times where I have something to formulated in my head to share to authors but never actually get around to doing it. I just gotta write it down and press send" (Quintana, 2025).

To that, we reply that tapping into the file drawer of informal peer reviews may be possible with one-click posting from reference managers and digital notebooks to PubPeer and the comments sections of preprint servers. To our knowledge, no existing technologies permit this. Similarly, one might enable users to crosspost their social media posts as PubPeer comments, like the Twitter account @PubPeerbot used to do. Meeting users in their digital bazaars and integrating crowdsourcing mechanisms, we hypothesize, promises greater scale than building more competing megachurches. Infrastructural elements that rest upon other infrastructure to connect people and tools are *glueware* – simple, low-cost, and productive means to connect infrastructural elements.

More radically, we might solicit more informal peer reviews by building a bespoke, integrated Publish-Review-Curate platform wherein authors preprint, other users review, and journals curate the reviewed preprints into collections for readers. This ideal is materializing quickly, as decentralized preprint servers like The Chive (chive.pub) allow scholars to upload their preprints and request "community peer review" from interested volunteers.

Just as authoring informal peer reviews is difficult, so too is *discovering and aggregating* informal peer reviews for rapid sensemaking. Researchers have expressed this information need as a particularly salient pain point when asking for "... people expert in research methods [to] publicise lists of research they have assessed as adequate and at least potentially significant... (Berkley, 2026b) "Journal editors are also beleaguered sorting through scattered informal peer reviews to the point that the editor of *British Medical Journal* now advises "readers and authors to use our rapid response service, as we're unable to keep track of criticism that can appear in many places online or on social media - and we make no promise to do so. (Abbassi, 2025)" Not all journals turn a blind eye to informal reviews off-site. Others like Frontiers and PLOS monitor PubPeer and respond to social media posts. When informal and formal reviews (as letters) alert Taylor & Francis to problematic papers, they post notices on their webpages about articles flagged with serious concerns but not yet a formal Correction or Notice of Editorial Concern. Informal peer reviewers have remarked that these peripheral notices are not visible enough for them and seem disconnected from copies of articles readers encounter away from publishers' websites (Kincaid, 2024). Readers need a summary of informal peer reviews for a given scholarly report.

It will be valuable, as a first step, to connect the informal peer reviews *within* a social media site to other users who may find such reviews helpful, extending the Twitter Community Notes feature that corrects erroneous posts to Bluesky. Science journalists, when disseminating emerging research, may find informal peer reviews useful as they write public-facing articles in news media. One recently launched browser extension, AuthentiSci (authentisci.com), promises related value by enabling its users to nominate a media article for review from ORCID-validated researchers in a relevant field (AuthentiSci, 2024). Nominated researchers can score the strength of the evidence-conclusions connections, objectivity (balance), and clarity. Total scores and variability across reviewers are computed and visualized in a small widget. Glueware also supports users' just-in-time information acquisition by connecting PubPeer comments to journal articles and preprint webpages and between PubPeer pages and PDFs in reference managers. Together, a patchwork of glueware, small and large, can channel arguments and counterarguments across the information network to users in need.

From the start of our ethnography, we witnessed the flowering of technologies that *aggregate* informal peer reviews. In one prototyped paper-reading application, Surf, users reading a research report can also reference a summary of social media posts mentioning the report in a sidebar (Huang et al., 2025). According to self-reports, this just-in-time information was valuable to readers. Likewise, users on the preprint commentary platform alphaXiv (alphaxiv.org), which provides a social annotation layer over PDFs of arXiv preprints, are able to view on-platform and unsummarized comments of the report they're viewing if Twitter users commented (alphaXiv, 2025).

Tools like the browser extension LazyScholar (lazyscholar.org) also aggregate information about research reports, spanning PubPeer, the social annotation website Hypothesis, the informal review dashboard Problematic Paper Screener, and journal-provided bibliometrics (Vorland, 2026). A smaller, but more usable tool called RetractionRisk Scanner (retractionrisk.com) is likewise helpful for scanning across the web for positive and negative remarks about a scholarly report (Zheng, 2025). When fed references or when visiting a journal article webpage, it connects users to PubPeer comments, negative sentiment posts on Bluesky and Twitter, as well as notices of retraction status Similar dashboard-focused web apps like Signals (research-signals.com) aggregate risk factors for retractions using informal peer reviews from vetted experts, self-citations, and other signals in a single commercial web app (Research Signals Limited, 2024). Scholars interested in authoring and viewing informal peer review are no longer devoid of options; they can access ready-to-use tools and examples that inspire offshoot tools. The concomitant rise of problematic papers and AI technologies to detect them has, it seems, inspired many to develop technologies for discovering and aggregating informal peer reviews. See Table 6 for an overview of informal peer review platforms, glueware, and aggregators.

**Table 6.** Informal peer review platforms include bespoke platforms, social media, and preprint servers. Glueware and aggregators link data between platforms.

| # | Name | URL | Participation | Interaction | Platform |
|---|---|---|---|---|---|
| *Bespoke platforms* | | | | | |
| 1 | PubPeer | pubpeer.com | anyone | moderated | free |
| 2 | ResearchHub | researchhub.com | anyone | unmoderated | free |
| 3 | alphaXiv | alphaXiv.org | anyone | unmoderated | free |
| 4 | Paperstars | paperstars.org | anyone | moderated | free |
| *Social media* | | | | | |
| 1 | Bluesky | bsky.app | anyone | unmoderated | free |
| 2 | Twitter / X | x.com | anyone | unmoderated | freemium |
| 3 | Reddit | reddit.com | anyone | unmoderated | free |
| 4 | LinkedIn | linkedin.com | anyone | unmoderated | freemium |
| *Preprint servers* | | | | | |
| 1 | bioRxiv | bioarxiv.org | anyone | unmoderated | free |
| 2 | medRxiv | medRxiv.org | anyone | unmoderated | free |
| 3 | Chive | chive.pub | anyone | unmoderated | free |
| *Glueware* | | | | | |
| 1 | PubPeer browser ext. | pubpeer.com/static/extensions | NA | NA | free |
| 2 | PubPeer Zotero ext. | github.com/PubPeerFoundation | NA | NA | free |
| 3 | AuthentiSci browser ext. | authentisci.com | anyone | moderated | free |
| 4 | @PubPeerbot [inactive] | x.com/PubPeerBot | anyone | unmoderated | free |
| *Aggregators* | | | | | |
| 1 | RetractionRisk Scanner | retractionrisk.com | NA | NA | free |
| 2 | Lazy Scholar | lazyscholar.org | NA | NA | free |
| 3 | Signals [also solicits] | research-signals.com | anyone | moderated | freemium |
| 4 | Surf [inactive prototype] | 10.1145/3746059.3747647 | NA | NA | NA |

Each of our four enlightening themes offers guidance for scaling informal peer review. T1: Motley Crew suggests that widening the circle of evaluators beyond credentialed, active researchers will generate many diverse contributions rapidly and that many of these contributions will be uniquely useful. More radically, we contend that formal peer review should also adopt such inclusivity toward reviewers. In the not-too-distant future, bots and agentic AI models may coordinate with humans and other AI agents as peers.

T2: Hacking Homes by Nesting and Bridging suggests that developing communities on existing, active platforms will be fruitful, especially when paired with growth interventions. One might begin to post regularly on some topic of interest, develop groups on LinkedIn (now a popular hub for academic researchers, practitioners, and interested laypeople), and grow existing subreddits. In the nascent ATProtocol Science community, leveraging the ATProtocol tech infrastructure powering Bluesky, scholar-developers are building a Bluesky client, Lea, to assist researchers with safety, moderation, and the discovery of discussions about papers and preprints (Antoniak, 2026). In time, this nest and others like it may surface the most relevant informal peer reviews for the right user at the right time. They should be bridged within and across platforms using available and custom-made tools. Developing bots, in the spirit of @PubPeerbot, will also be productive. Beyond this, databases and knowledge hubs that automatically synthesize informal peer review discourse across platforms and with the consent of users may one day be mainstream and taken-for-granted. Readers may consult a unified, trusted resource resembling Yelp.

If a defined home with agreeable rules, technical affordances, and incentives for informal peer review were available, then the X-odus would have been less socially destructive. With luck, patchwork evidence infrastructures will not be needed long-term. Bespoke platforms like The Chive are gaining ground. Stable and defined homes are being constructed where authors and open-access reports nestle alongside commenters. This connection to authors near a familiar platform may help legitimate their invisible labor and elevate its status in the landscape of scholarly knowledge work. We predict that general-purpose and bespoke nests will codevelop and define the future of this patchwork.

T3: Intensive Infrastructural Inversion suggests that for informal and formal peer review to make substantive points, it will be useful to learn forensic metascience techniques and tools that make their routine application bearable. In our estimation, these techniques are little-known outside of metascientific circles. We cannot assume that the traditional techniques of reviewing will be sufficient for unearthing issues or that more modern techniques will permeate quickly enough to be applied. Greater investment in training materials like Collection of Open Science Integrity Guides (COSIG), the *de facto* operating manual for informal peer reviewers, may be wise (Richardson, 2026).

T4: Immunologic Self-preservation of the authors-editors-publishers triad teaches us that silence does not imply excellence post-publication. Publishers' minimally corrective measures are insufficient to communicate the credibility of scholarship. We cannot rely on concentrated powers like publishers, whether they are elite, mid-tier, or low tier. The power and legitimacy to regulate scholarship should be distributed throughout the scholarly community, where scholars are increasingly building their own infrastructure spanning methods, technologies, niches, and publication venues.

Beyond solving specific information needs with technologies and community-building, we can also reflect on two deep principles to structure research evaluation more generally. We find them to be the most consequential implications. First, we pay respects to *agency*. As a required and often opaque practice, formal peer review does not by its nature instill much agency in authors. This need not be the case. We might seek inspiration from book authors in the humanities who supplemented the traditional book review process with their own self-led review process, soliciting feedback from colleagues and 'clever friends' (Butchard et al., 2018). Their informal or "DIY peer review" exemplifies a focus on quality control missing in the practice of formal peer review, where authors' central goal is not self-improvement, but publication. No matter the eventual course of peer review, formal or informal, we would be prudent take charge and engage with all options to enhance our weakest work.

Platforms for reviewing research should imbue agency in users. When responding to my question on preferred platforms for informal review, two persistent informal peer reviewers and established academics reported that they preferred general-purpose social media sites like Bluesky over bespoke forums like PubPeer for their viral reach, editing capabilities (users cannot delete their own PubPeer posts), and casual conversations. On Twitter, another persistent informal reviewer remarked that PubPeer's aggressive intellectual property licensing of users' posts gave him pause. Just as hacker culture prioritizes openness, inclusivity, innovation, informality, and self-organized collaboration, informal review culture treasures the same values. Tool developers and community organizers, we surmise, may succeed in developing bespoke reviewing platforms if they center on scholars' values rather than launch a product disconnected from them.[46] One should "Move slowly and build things" (Abrams, 2025).

In addition to agency, we favor the principle of *agility*. Authors should organize their peer reviews, formal and informal, around speed and long-term maintenance. This requires that reviews be public, partitioned, and perpetual. Consider your next research project from its start. As you proceed through the research process, you may wish to expand the circle of evaluators and request feedback on specific components of the research from those who may add value. These activities would be ideal to implement before data sourcing to act on the feedback easily. Partitioning the review into a list of desired subtasks that specific types of reviewers could perform would eradicate the perennial "Now what do I do?" problem that many users face on informal peer review platforms. Structuring reviews to be public and partitioned, we think, implies that they will be perpetual. A solitary comment on a snippet of text may consistently accrue attention over years and decades such that the digital marginalia resemble a professionally annotated classical text.

## 7.5 Gaps & Future Research

Opportunities to engineer solutions and implement research abound. We think that the constructive, or tech infrastructure gaps, are especially vital to close. One might build the tools that users lack using the AT Protocol or propose hackathons to ship the bots, aggregation services, and glueware to connect informal peer reviewers to their beneficiaries. Such technology ought to be more plentiful and stably funded. As one reads a science news article, why can a browser extension not overlay a summary of the critical discourse atop the webpage for the layperson to use? Why do we not have an alphaXiv for psyarXiv, bioRxiv, and medRxiv? What about commentary on preregistration platforms? Where are the dependable, low-friction tools to unite sleuths and would-be sleuths on Bluesky, Twitter, LinkedIn, and other platforms?

---

[46] We believe that the merits of the decentralized ATProtocol are best aligned with scholars' values and leveraging this technical infrastructure will increase the chances of growing a sizeable userbase.

How are we to glean the informal peer reviews across the many venues where they happen? Conferences, private Slack groups, journal clubs, e-mail threads, and courses are home to stimulating discussions that evaporate or stagnate in digital dustbins. When authors receive feedback on their bioRxiv preprint, the plurality (36%) get it by e-mail, 34% get it through in-person conversations, 32% through Twitter, and just 9% through bioRxiv comments (Sever et al., 2026). A majority of survey respondents reported preferring email for preprint feedback. The dark matter of private informal peer reviews and even neutral to positive comments are worth illuminating. Future efforts ought to consider how one might do that as well as open the file drawer of unspoken, unwritten reviews. Perhaps it might involve leveraging the sequential strategy of change successfully used by the Center for Open Science: make it possible, make it easy, make it normative, make it rewarding, and make it required (Nosek, 2019).

Research, in the near future, can more meticulously work out a detailed ontology of the family of peer review variants, defining and relating them. Theoreticians might explore how to best govern informal peer review and the optimal balance between open participation and closed work. They may ask and answer the puzzle of how legitimacy operates within formal and informal peer review. Whatever one’s ontic inclinations or theoretic proclivities, though, empirical researchers are advised to study each form of peer review at eye-level, following many cases across myriad settings and over a time span long enough to capture its intricate identity.

## 7.6 Conclusion

When this study was not yet conceived, I (JP) advised a longtime mentor to join Academic Twitter. His quizzical look and skeptical reply exposed entrenched assumptions about where deep work can brew. Scholarly rituals like formal peer review, as the cyclical complaints levied against it suggest, have not earned their monopoly on research evaluation. They are cultural fixations whose assumptions have not been met. In our digital ethnography, we found that the alternative, informal peer review, is eyed suspiciously and when poorly conceived initiatives and pet projects fail, its standing falls incommensurately lower. As one thoughtful voice cautioned, "The stars have begun to align for open [informal] peer review but fall short of a full constellation (Boston, 2024)." Here, we have argued that diverse and detailed descriptions of informal review's operations and immersion in its culture trod paths toward that "full constellation." Those paths, which oppose the failed frames of past reformers, call for investment in agency, agility, informality, and interoperability.

For our part, we are investing in these principles here by soliciting *your* engagement directly. As we shift from authoring this report to finalizing it for publication, we invite you to leave your most honest, unbridled, and sweeping informal peer review. Remark on a loose argument, a shaky citation, or a missed opportunity to connect to a related matter unknown to us. Request clarifying comments, whether in your first reading or years afterward. Counter that a stated contribution is not quite as novel as we suppose in a lengthy, citation-packed passage or in the terse, choppy fragments of an unfinished argument. Question the societal implications of researching and practicing informal peer review whether or not you fully know what they fully entail. In short, take to heart the most incontrovertible lesson that informal peer reviewers teach – that you can just review things.

## Contributors

We are grateful to Susannah Paletz for lending a lab room as well as volunteers Amelia Gibson, Irene Pasquetto, Zubin Jelveh, Giovanni Luca Ciampaglia, and Michael Mahon for critical feedback on early versions of this report (Table 7).

**Table 7.** Contributor roles including co-authors (JP, JC) and volunteers (IP, SP, ZJ, GC, AG, MM).

| Role | Co-authors | | Volunteers | | | | | |
|---|---|---|---|---|---|---|---|---|
| | JP | JC | IP | SP | ZJ | GC | AG | MM |
| Conceptualization | high | high | none | none | none | none | none | none |
| Data curation | high | low | none | none | none | none | none | none |
| Formal analysis | high | high | none | none | none | none | none | none |
| Funding acquisition | NA | NA | NA | NA | NA | NA | NA | NA |
| Investigation | high | high | none | none | none | none | none | none |
| Methodology | high | high | medium | none | none | none | none | none |
| Project administration | high | high | none | none | none | none | none | none |
| Resources | high | high | none | high | none | none | none | none |
| Software | high | high | none | none | none | none | none | none |
| Supervision | high | high | mod | none | none | none | none | none |
| Validation | high | high | none | none | none | none | none | none |
| Visualization | high | none | none | none | none | none | none | none |
| Writing – Original Draft | high | low | none | none | none | none | none | none |
| Writing – Review & Editing | high | high | high | none | high | high | high | low |

## AI Use Statement

We used each tool with full human supervision and accept responsibility for any errors.

JP: ChatGPT Images generated the cover image. I did not use any AI tools to ideate questions, design the study, source data, analyze, or interpret any insights here. I did supplement my database searches for relevant publications with AI-enhanced search engines including Elicit (www.elicit.org) and Undermind AI (www.undermind.ai). I also used Grammarly to proofread my text and LLMs (Claude 4, Gemini 2.5 Pro, and ChatGPT o3 pro) for a range of checks after authoring the first draft myself. In later stages, I used Claude Opus 4.6, GPT-5.4 Thinking, Gemini 3.1 Pro, and DeepSeek R1 (DeepThink) to review and proofread the manuscript, catching many errors.

JC: Nothing to disclose for this project.

## Funding

None

## Competing Interests

JP: See my Positionality and Reflexivity Statement above in §2 above for a detailed description of my relation to peer review. I acknowledge participation in reform peer reviewing initiatives like ASAPBio preprint club – Meta-research (unpaid, 2025 onward) and RepliCATS (two workshops, paid small gifts). These initiatives share some features of informal peer review, but are structured with administrators and prespecified operating procedures. I have informally advised and participated in events with early-stage startups like alphaXiv, PaperStars, semble (https://semble.so/), and am now volunteering in the Continuous Review Foundation's Modular Peer Review Working Group (unpaid).

JC: I acknowledge participation in digital scholarly infrastructure reform efforts, both informally, via networks of initiatives like ATProto Science and the Continuous Science Foundation, and formally, via direct facilitation and participation in workshops like the Navigation Fund funded workshop on Catalyzing Modular Interoperable Research Attribution. My interest is in unbundling of peer review into more modular research objects that support more granular scholarly collective sensemaking, which I am working towards via my co-direction of the Discourse Graphs project (https://discoursegraphs.com/).

## Data Availability Statement

Our open dataset and codebook are available on GitHub at https://github.com/oasisresearchlab/informal-peer-review and Figshare at https://doi.org/10.6084/m9.figshare.33187710.